\documentclass[aps,prd,amsmath,amssymb,twocolumn,superscriptaddress,preprintnumbers,nofootinbib]{revtex4-1}

\usepackage[utf8]{inputenc}
\usepackage{graphicx}
\usepackage{dcolumn}
\usepackage{bm}
\usepackage[bookmarks, pagebackref=false]{hyperref}
\usepackage[nameinlink]{cleveref}
\usepackage[usenames,dvipsnames]{xcolor}
\usepackage[normalem]{ulem}
\definecolor{rossoCP3}{cmyk}{0,.88,.77,.40}
\definecolor{blaa}{rgb}{0.2,0.2,0.6}
\hypersetup{colorlinks,
	bookmarksopen,
	bookmarksnumbered,
	citecolor=blaa, 		
	linkcolor=rossoCP3,	
	urlcolor=rossoCP3,			
}

\crefname{section}{Sec.\!}{Secs.\!}
\crefname{figure}{Fig.\!}{Figs.\!}
\crefname{equation}{}{}
\crefname{table}{Tab.\!}{Tabs.\!}
\crefname{appendix}{App.\!}{Apps.\!}

\usepackage{natbib}
\usepackage{slashed}
\newcolumntype{x}[1]{>{\centering\arraybackslash\hspace{0pt}}p{#1}}
\usepackage{makecell}

\def\MeV{\,\text{MeV}}
\def\GeV{\,\text{GeV}}
\def\TeV{\,\text{TeV}}
\def\Tr{~{\mbox{Tr}}}
\def\tr{~{\mbox{tr}}}
\def\mL{\mathcal{L}}
\def\Gnob{G}
\def\signob{\sigma}
\def\Gbar{\bar{G}}
\def\sigbar{\bar{\sigma}}

\graphicspath{{./figures/}}

\begin{document}

\title{ \LARGE  \color{rossoCP3} Dark Confinement and Chiral Phase Transitions: Gravitational Waves vs Matter Representations}

\author{Manuel {\sc Reichert}}
\thanks{{\scriptsize Email}: \href{mailto:m.reichert@sussex.ac.uk}{m.reichert@sussex.ac.uk}; {\scriptsize ORCID}: \href{https://orcid.org/0000-0003-0736-5726}{ 0000-0003-0736-5726}}
\affiliation{Department  of  Physics  and  Astronomy,  University  of  Sussex,  Brighton,  BN1  9QH,  U.K.}

\author{Francesco {\sc Sannino}}
\thanks{{\scriptsize Email}: \href{mailto:sannino@cp3.sdu.dk}{sannino@cp3.sdu.dk}; {\scriptsize ORCID}: \href{https://orcid.org/0000-0003-2361-5326}{ 0000-0003-2361-5326}}
\affiliation{Scuola Superiore Meridionale, Largo S. Marcellino, 10, 80138 Napoli NA, Italy}
\affiliation{CP$^3$-Origins,  University of Southern Denmark, Campusvej 55, 5230 Odense M, Denmark}
\affiliation{Dipartimento di Fisica “E. Pancini”, Università di Napoli Federico II | INFN sezione di Napoli, Complesso Universitario di Monte S. Angelo Edificio 6, via Cintia, 80126 Napoli, Italy}

\author{Zhi-Wei {\sc Wang}}
\thanks{{\scriptsize Email}: \href{mailto:wang@cp3.sdu.dk}{wang@cp3.sdu.dk}; {\scriptsize ORCID}: \href{https://orcid.org/0000-0002-5602-6897}{0000-0002-5602-6897}}
\affiliation{CP$^3$-Origins,  University of Southern Denmark, Campusvej 55, 5230 Odense M, Denmark}
\affiliation{Department of Astronomy and Theoretical Physics, Lund University, 22100 Lund, Sweden}

\author{Chen {\sc Zhang}}
\thanks{{\scriptsize Email}: \href{mailto:chen.zhang@fi.infn.it}{chen.zhang@fi.infn.it}; {\scriptsize ORCID}: \href{https://orcid.org/0000-0003-2649-8508}{0000-0003-2649-8508}}
\affiliation{INFN Sezione di Firenze, Via G. Sansone 1, I-50019 Sesto Fiorentino, Italy}

\begin{abstract}
We study the gravitational-wave signal stemming from strongly coupled models featuring both, dark chiral and confinement phase transitions. We therefore identify strongly coupled theories that can feature a first-order phase transition. Employing the Polyakov-Nambu-Jona-Lasinio model, we focus our attention on $SU(3)$ Yang-Mills theories featuring fermions in fundamental, adjoint, and two-index symmetric representations. We discover that for the gravitational-wave signals analysis, there are significant differences between the various representations. Interestingly we also observe that the two-index symmetric representation  leads to the strongest first-order phase transition and therefore to a higher chance of being detected by the Big Bang Observer experiment. Our study of the confinement and chiral phase transitions is further applicable to extensions of the Standard Model featuring composite dynamics.
\end{abstract}

\maketitle

\section{Introduction}
Future gravitational-wave (GW) observatories provide new opportunities to investigate the existence of dark sectors that are currently inaccessible. Because the experiments will be sensitive to strong first-order phase transitions in the early universe, it is paramount to understand the landscape of theories that can lead to GW signals. Within the landscape of theories, asymptotically free gauge-fermion systems are privileged, being already chosen by nature to constitute the backbone of the Standard Model (SM). It is therefore reasonable to expect them also to appear in the dark sector \cite{Kumar:2011iy, DelNobile:2011je, Hietanen:2013fya, Boddy:2014yra, Boddy:2014qxa, Hochberg:2014dra, Hochberg:2014kqa, Cline:2016nab, Cacciapaglia:2020kgq, Dondi:2019olm, Ge:2019voa, Beylin:2019gtw, Yamanaka:2019yek, Yamanaka:2019aeq}.

In \cite{Huang:2020crf}, we embarked on a systematic investigation of the composite landscape by providing the first comprehensive study of the dark confinement phase transition stemming from pure gluonic theories. There, we first extended the state-of-the-art knowledge of the confinement phase transition to arbitrary number of colours by combining effective approaches with lattice results and then determined their GW imprints.

Using our work \cite{Huang:2020crf} on pure gauge dynamics as a stepping stone, we now investigate the dark dynamics stemming from the confinement and chiral phase transitions arising when adding fermions in different matter representations to gauge theories. For previous works and complementary approaches on chiral and confinement phase transitions see e.g.~\cite{Jarvinen:2009mh, Schwaller:2015tja, Chen:2017cyc, Helmboldt:2019pan, Agashe:2019lhy, Bigazzi:2020avc, Halverson:2020xpg,Kang:2021epo,Garcia-Bellido:2021zgu, Ares:2020lbt, Zhu:2021vkj, Ares:2021ntv, Li:2021qer}. We highlight below the main findings of our work:
\begin{table*}[t!]
	\resizebox{\textwidth}{15mm}{
	\begin{tabular}{|c|c|c|c|c|c|c|c|}
		\hline
		\;Model name\; & \;Gauge group\; & \makecell[c]{\;Fermion\; \\ irrep} & \;Reality\; & \;$N_f$\; & \;KMT term\; & \makecell[c]{Centre \\ \;symmetry\;} & \;Chiral symmetry breaking pattern\;\\
		\hline
		3F3 & $SU(3)$ & $\textbf{F}$ & C & 3 & 6F & $\varnothing$ &$SU(3)_L\times SU(3)_R\rightarrow SU(3)_V$ \\
		\hline
		3G1 & $SU(3)$ & $\textbf{Adj}$ & R & 1 & \makecell[c]{12F \\ (Ignored)} & $Z_3$ & $SU(2)\times U(1)_A\rightarrow SO(2) $$^\ast$ \\
		\hline
		3S1 & $SU(3)$ & $\textbf{S}_2$ & C & 1 & \makecell[c]{10F \\ (Ignored)} & $\varnothing$ &$U(1)_L\times U(1)_R\rightarrow U(1)_V$$^\ast$ \\
		\hline
	\end{tabular}}
	\caption{\label{tab:3models}
		The three dark gauge-fermion models studied here, see \cref{subsec:bc} for a detailed explanation.
	}
\end{table*}
\begin{enumerate}
	\item We have systematically investigated the chiral and confinement phase transitions including their interplay for $SU(3)$ Yang-Mills theory with fermions in fundamental, adjoint, and two-index symmetric representations via the Polyakov-Nambu-Jona-Lasinio (PNJL) model.
	\item Representations matter: fermions in the two-index symmetric representation increase the strength of the first-order confinement phase transition while the fermions in the adjoint representation decrease it. Notably, for the adjoint and two-index symmetric representation cases with one Dirac flavour, the chiral phase transition is a second-order phase transition.
	\item For all representations (chiral and confinement phase transitions), the inverse duration of the phase transition is large, $\tilde \beta\sim\mathcal{O}(10^4)$. We discuss the large value of $\tilde\beta$ in the thin-wall approximation where it is given by a competition of the surface tension with the latent heat. This  sheds light on generic features of non-abelian gauge-fermion systems and helps finding models with stronger gravitational-wave signals.
	\item The confinement phase transition with fermions in the two-index symmetric representation has the best  chance of being detected by the Big Bang Observer (BBO) with a potential signal-to-noise ratio $\text{SNR}\sim \mathcal{O}(10)$.
	\item We further provide an outlook of which strongly coupled models can potentially lead to a first-order chiral phase transition. The methods used in this work can be readily applied to some of these models as well as other dark and bright extensions of the Standard Model featuring new composite dynamics.
\end{enumerate}

This work is structured as follows. In \cref{sec:eff-theories}, we introduce the PNJL model as an effective low energy model which we use here to describe the dynamics in the dark gauge-fermion sector. In \cref{sec:phase-transition}, we discuss the details and interplay of the chiral and confinement phase transitions, including the bubble nucleation and the GW parameters and spectrum. In \cref{sec:results}, we present our results which includes a detailed analysis of the GW spectrum for the various models and the corresponding signal-to-noise ratios. In \cref{sec:discussion}, we discuss our results in the light of the thin-wall approximation and analyse which further models are likely to have a first-order phase transition. In \cref{sec:conclusions}, we offer our conclusions.

\section{Effective Theories for Dark Gauge-Fermion Sectors}
\label{sec:eff-theories}

\subsection{Basic Considerations}
\label{subsec:bc}
A first-order phase transition in the early universe generates a GW signal that might be detectable by future observations~\cite{Bian:2019bsn, Huang:2017rzf, Basler:2016obg, Dorsch:2017nza, Bernon:2017jgv, Gorda:2018hvi, Brdar:2019fur, Huang:2020bbe, Croon:2018kqn, Xie:2020bkl, Eichhorn:2020upj, Wang:2020jrd}, for recent reviews see, e.g., \cite{Caprini:2019egz, Hindmarsh:2020hop}. This offers unprecedented detection possibilities for dark sectors that are otherwise (almost) decoupled from the visible SM sector. In this section, we concentrate on the description of such dark phase transitions, which fall into two categories:
\begin{enumerate}
	\item \textbf{Confinement phase transition} characterized by the restoration of the centre symmetry at low temperatures~\cite{McLerran:1981pb}. The order parameter is the traced Polyakov loop~\cite{Polyakov:1975rs}.
	\item \textbf{Chiral phase transition} characterized by the spontaneous breaking of chiral symmetry at low temperatures. The order parameter is the chiral condensate.
\end{enumerate}
These phenomena occur in the strongly coupled regime of the gauge dynamics and cannot be described by perturbation theory. To make further progress, three approaches can be envisioned:
\begin{enumerate}
	\item Universality analysis~\cite{Svetitsky:1982gs, Pisarski:1983ms, Svetitsky:1985ye}. This is useful for investigating the order of the phase transition but does not provide a quantitative way to compute thermodynamic observables near a first-order phase transition.
	\item First-principle non-perturbative approaches, e.g.\ lattice gauge theory~\cite{Karsch:2001cy} and functional renormalization group~\cite{Fu:2019hdw}. Unlike pure-gauge theories, first-principle results for thermodynamic observables are very limited for gauge-fermion theories in the chiral limit (i.e.\ zero fermion mass limit).
	\item Effective theories~\cite{Fukushima:2013rx, Fukushima:2017csk}. These are constructed by including the relevant degrees of freedom and enforcing symmetry principles. They allow for a quantitative framework to compute thermodynamic observables near a first-order phase transition. However, there is the possibility that they do not provide a faithful modelling of the phase transition dynamics. The results obtained from effective theories provide valuable hints about possible dynamics of the underlying gauge-fermion theory, but should always be interpreted with care.
\end{enumerate}
In this work, we employ effective theories to obtain a quantitative description of the phase-transition dynamics in dark gauge-fermion sectors. For the description of the chiral phase transition at finite temperature, chiral effective theories such as the Nambu-Jona-Lasinio (NJL) model~\cite{Nambu:1961tp, Nambu:1961fr}, see \cite{Vogl:1991qt, Klevansky:1992qe, Hatsuda:1994pi, Buballa:2003qv} for reviews, and the Quark-Meson (QM) model~\cite{Ellwanger:1994wy, Jungnickel:1995fp} are frequently adopted in the literature. In order to account also for the confinement phase transition, these models have been generalized to include the Polyakov-loop dynamics, with quarks propagating in a constant temporal background gauge field. The resulting models are called Polyakov-Nambu-Jona-Lasinio (PNJL) model~\cite{Fukushima:2003fw, Ratti:2005jh} and Polyakov-Quark-Meson (PQM) model~\cite{Schaefer:2007pw, Skokov:2010sf}, respectively. See also \cite{Mocsy:2003qw} for a study of the Polyakov-extended linear sigma model. Here, we focus on the PNJL approach and leave the PQM approach for future work.

We study three dark gauge-fermion models as shown in \cref{tab:3models}. In the ``Fermion irrep" column, $\textbf{F}$, $\textbf{Adj}$, and $\textbf{S}_2$ denotes the fundamental, adjoint, and two-index symmetric representation, respectively. The ``Reality" column refers to the reality property of the fermion representation, which is complex (C) for $\textbf{F}$ and $\textbf{S}_2$, and real (R) for $\textbf{Adj}$. The ``KMT term" column indicates the number of fermions needed to form a Kobayashi-Maskawa-'t Hooft term~\cite{Kobayashi:1970ji, tHooft:1976rip} (i.e.\ the 't Hooft determinantal term), e.g.\ 6F means the KMT term is a six-fermion interaction. For the 3G1 and 3S1 models the KMT terms are 12-fermion and 10-fermion terms respectively, and in our PNJL treatment, their effects are ignored since they correspond to operators of very high dimensions. The chiral symmetry breaking patterns for 3G1 and 3S1 models are also marked with an asterisk to indicate that the effects of KMT term are not considered (so the $U(1)_A$ part is included in the chiral symmetry).

We restrict our attention to the $SU(3)$ gauge group for simplicity. The treatment of the PNJL grand potential in gauge groups of larger rank requires introducing two or more independent Polyakov-loop variables and is left for future work. Three smallest fermion representations $\textbf{F}$, $\textbf{Adj}$, and $\textbf{S}_2$ are then the natural targets for investigation. For the fundamental representation case, we consider the three Dirac flavour case, which is most likely to exhibit a first-order chiral phase transition (see discussion on universality below). The phase transition and GW signature of the 3F3 model have been also studied in \cite{Helmboldt:2019pan}, using several effective theories including the PNJL model. Compared to \cite{Helmboldt:2019pan}, we employ a different regularisation as discussed below. For the adjoint and two-index symmetric representations, we consider one Dirac flavour since the case of two Dirac flavours is believed to be close to the lower boundary of the conformal window~\cite{DelDebbio:2009fd,Sannino:2009za,Rantaharju:2015cne}\footnote{If the two Dirac flavour case is inside the conformal window, then there is no confinement and chiral symmetry breaking at low temperature. It is also possible that the two Dirac flavour case is just below the lower boundary of the conformal window, associated with a weakly first-order phase transition in the sense discussed in \cite{Gorbenko:2018ncu}. Some nice discussions also appear in \cite{Tuominen:2012qu}.}. We work in the chiral limit, i.e.\ setting all current quark masses to zero for simplicity. In a more generic setup, nonzero current quark masses could be introduced in such cases, leading to pseudo-Nambu-Goldstone bosons (pNGB). We leave this more generic case for future study.

\subsection{The PNJL Model}
\label{subsec:2B}
The PNJL Lagrangian can be generically written as~\cite{Fukushima:2003fw, Ratti:2005jh}
\begin{align}
	\mL_{\rm{PNJL}}=\mL_{\text{pure-gauge}}+\mL_{\rm{4F}}+\mL_{\rm{6F}}+\mL_k \,,
\end{align}
where $\mL_{\text{pure-gauge}}$ is the pure-gauge part of the Lagrangian whose effect is to contribute as
the Polyakov-loop potential in the full grand potential, to be described below. $\mL_{\rm{4F}}$ and $\mL_{\rm{6F}}$
are the multi-fermion interaction terms which exist in the NJL model. They read
\begin{widetext}
	\begin{align}
		\mathrm{3F3:}\,
		&\mL_{\mathrm{4F}}=\Gnob_S\sum_{a=0}^8[(\bar{\psi}\lambda^a\psi)^2+(\bar{\psi}i\gamma^5\lambda^a\psi)^2]\,,
		&\mL_{\mathrm{6F}}=&\Gnob_D[\mathrm{det}(\bar{\psi}_{Li}\psi_{Rj})+\mathrm{det}(\bar{\psi}_{Ri}\psi_{Lj})] \,, \label{eq:3F3L} \\
		\mathrm{3G1:}\,
		&\mL_{\mathrm{4F}}=\Gnob_S\Big[(\bar{\psi}\psi)^2+(\bar{\psi}i\gamma^5\psi)^2
		+|\overline{\psi^C}P^*\psi|^2+|\overline{\psi^C}P^*\gamma^5\psi|^2 \Big],
		&\mL_{\mathrm{6F}}=&0\,, \label{eq:3G1L} \\
		\mathrm{3S1:}\,
		&\mL_{\mathrm{4F}}=\Gnob_S(\bar{\psi}_L\psi_R)(\bar{\psi}_R\psi_L)\,,
		&\mL_{\mathrm{6F}}=&0\,, \label{eq:3S1L}
	\end{align}
\end{widetext}
Here the same notation $\psi$ for the dark quark fields, and $\Gnob_S$ for the four-fermion coupling are adopted in all three models to avoid proliferation of new symbols. Note that $\psi$ is a colour triplet in 3F3, a colour octet in 3G1, and a colour sextet in 3S1. It is understood that in the above equations, the colour indices of $\psi$ are contracted to form singlets inside the fermion bilinear it resides in. For example, $\bar{\psi}$ and $\psi$ are contracted with a Kronecker delta in colour space, while $\overline{\psi^C}$ and $\psi$ are contracted with a unitary symmetric matrix $P^*$ in the colour space in the 3G1 case\footnote{$\psi^C$ is defined as $\psi^C\equiv C\bar{\psi}^T$ as usual, with $C$ being the charge conjugation matrix. The existence of the unitary symmetric matrix $P$ is guaranteed in the 3G1 case because the adjoint representation is real.}. In the 3F3 model case, $\psi$ is also a three-component vector in the flavour space and we write $\psi=(u,d,s)^T$ when we want to make its individual components explicit. Note that these are dark quarks, not to be confused with the SM quarks. The $\lambda^a$ are $3\times 3$ matrices in flavour space with $\lambda^0\equiv\sqrt{\frac{2}{3}}\,\textbf{1}$ and $\lambda^a$, with $a=1,...,8$, are the usual Gell-Mann matrices written in the flavour space, normalized as $\text{Tr}_F(\lambda^a\lambda^b)=2\delta^{ab}$ ($\text{Tr}_F$ denotes trace in the flavour space only). Finally, $\mL_k$ is the covariant kinetic term for the quark field~\cite{Fukushima:2003fw,Ratti:2005jh}
\begin{align}
	\mL_k&=\bar{\psi}i\gamma_u D^\mu\psi\,,
	&
	D^\mu&=\partial^\mu-iA^\mu\,,
	\label{eq:ckt}
\end{align}
with $A^\mu=\delta_0^\mu A^0$ being the temporal background gauge field living in the corresponding fermion representation. In the Polyakov gauge~\cite{Weiss:1980rj, Fukushima:2003fw}, $A^0$ can be taken to be diagonal and static. In the mean-field approximation to be introduced later, $A^0$ is also taken to be spatially homogeneous so it acts as a constant imaginary chemical potential~\cite{Fukushima:2003fw} ($A^0=-iA_4$ when continued to Euclidean spacetime). This way of coupling the gauge field to the quark field allows us to investigate the interplay between confinement and chiral dynamics in a convenient manner. However, only temporal gauge fields play a role in the modelling here. It is expected that for high temperatures (a few times the confinement phase-transition temperature), the transverse gluons are also important and the PNJL modelling should be revised accordingly~\cite{Meisinger:2003id, Ratti:2005jh}.

A few remarks are in order regarding the construction of $\mL_{\rm{4F}}$ and $\mL_{\rm{6F}}$ in the NJL model. In principle one should write down to a given order (e.g.\ four-fermion level) all operators that are compatible with the full symmetry (spacetime, colour, and flavour) of the theory, with each independent operator carrying an independent coefficient~\cite{Bernard:1987sg, Klimt:1989pm}. A further complication arises due to the fact that in computing fermion loops, both the direct term and the exchange term may contribute. One can achieve simplifications by considering a Fierz-invariant Lagrangian from the beginning, which can be obtained by adding to the original Lagrangian its Fierz transformation~\cite{Klimt:1989pm, Klevansky:1992qe}. With the Fierz-invariant Lagrangian, one only needs to consider the direct terms in the computation~\cite{Klevansky:1992qe}. Fortunately, in many cases (including the present work) we do not need to carry out the exercise of writing down the full Fierz-invariant Lagrangian compatible with the symmetry of the theory. This is because one may work in a mean-field approximation and only care about the condensate in some but not all channels. For example in the mean-field approximation here, we are only concerned with the condensate $\langle\bar{\psi}\psi\rangle$ since we work at finite temperature but zero chemical potential. We therefore only need to retain four-fermion terms related to this particular channel and express them in a form that preserves the full symmetry of the theory. Adding the Fierz transformation amounts to a redefinition of the couplings in the terms relevant for calculation. Since the coupling $\Gnob_S$ and $\Gnob_D$ are left arbitrary at this stage, we may also stick to the Fierz-non-invariant Lagrangian and take into account only the direct terms in the computation, without loss of generality~\cite{Buballa:2003qv}.

The chiral symmetry of the $\mL_{\rm{4F}}$ term in the 3F3 case can be made manifest by introducing composite fields $\Phi_{ij}\equiv\bar{\psi}_{jR}\psi_{iL},\,
(\Phi^\dagger)_{ij}\equiv\bar{\psi}_{jL}\psi_{iR}$ with $i,j=1,2,3$ being flavour indices and making use of the following identity for the four-fermion term~\cite{Hatsuda:1994pi,Holthausen:2013ota}
\begin{align}
	\Tr_F(\Phi^\dagger\Phi)=\frac{1}{8}\sum_{a=0}^8\left[(\bar{\psi}\lambda^a\psi)^2+(\bar{\psi}i\gamma^5\lambda^a\psi)^2\right],
\end{align}
which can be proven by brute force using the explicit form of the $\lambda^a$ matrices. $\Phi$ transforms as $(\textbf{3},\bar{\textbf{3}})$ under the chiral symmetry group $SU(3)_L\times SU(3)_R$, and thus $\Tr_F(\Phi^\dagger\Phi)$ is invariant under $SU(3)_L\times SU(3)_R$. The six-fermion term $\mL_{\rm{6F}}$ in the 3F3 case is a parametrization of the $U(1)_A$-breaking instanton effect. It is also chirally-invariant because it can be written as $\Gnob_D(\mathrm{det}\Phi+\mathrm{h.c.})$ and the $SU(3)_L$ and $SU(3)_R$ transformation matrices have unit determinants. In the 3G1 model, the $\mL_{\rm{4F}}$ term can be rewritten using the Nambu spinor $\Psi\equiv\left(\begin{matrix} \psi_L \\ P\psi_R^C\end{matrix}\right)$ as~\cite{Zhang:2010kn}
\begin{align}
	\mL_{\mathrm{4F}}&=\Gnob_S\Big[(\bar{\psi}\psi)^2+(\bar{\psi}i\gamma^5\psi)^2 \notag \\
	&\hspace{1cm}+|\overline{\psi^C}P^*\psi|^2+|\overline{\psi^C}P^*\gamma^5\psi|^2 \Big] \notag \\
	&=\Gnob_S|\overline{\Psi^C}\vec{\Sigma}\Psi|^2\,,
\end{align}
in which $\Psi^C\equiv P\left(\begin{matrix} \psi_L^C \\ (P\psi_R^C)^C\end{matrix}\right)$ and the $2\times 2$ $\vec{\Sigma}$ matrices are defined by $\Sigma_1=\textbf{1},\, \Sigma_2=i\sigma_1,\,\Sigma_3=i\sigma_3$. The unitary matrix $P$ is a symmetric matrix in colour space such that $P\psi_R^C$ has the same gauge transformation properties under the gauge group as $\psi_L$. This rewriting of $\mL_{\rm{4F}}$ using Nambu spinors makes the $SU(2)\times U(1)_A$ symmetry manifest in the 3G1 case, since $\Psi$ transforms as a doublet in the Nambu space, while $\overline{\Psi^C}\vec{\Sigma}\Psi$ can be shown to transform as a complex three-vector in the Nambu space\footnote{In \cite{Zhang:2010kn} there is a second term written at the four-fermion level for the NJL Lagrangian of one Dirac flavour of quark in a real representation of the gauge group. That term explicitly breaks $U(1)_A$ (but not to a correct discrete subgroup) and is mistaken for the 't Hooft determinental interaction by the authors. We therefore do not include it.}. Finally, in the 3S1 case, the $\mL_{\rm{4F}}$ term in \cref{eq:3S1L} is already manifestly invariant under the chiral $U(1)_L\times U(1)_R$ symmetry.

The finite-temperature grand potential of the PNJL models can be generically written as
\begin{align}
	V_{\rm{PNJL}}&=V_{\rm{PLM}}[\ell,\ell^*]+V_{\rm{cond}}\!\left[\langle \bar{\psi}\psi\rangle\right] \notag \\ &\quad\,+V_{\rm{zero}}\!\left[\langle \bar{\psi}\psi\rangle\right]
	+V_{\rm{medium}}\!\left[\langle \bar{\psi}\psi\rangle, \ell,\ell^*\right].
	\label{eq:grandpotential}
\end{align}
Here $\ell$ and $\ell^*$ denote the traced Polyakov loop and its conjugate (to be defined more precisely below). $V_{\rm{PLM}}[\ell,\ell^*]$ is the Polyakov-loop potential describing the glue sector where all the potential can be fully determined by the existing lattice results (see e.g.~\cite{Huang:2020crf}). $V_{\rm{cond}}\!\left[\langle \bar{\psi}\psi\rangle\right]$ represents the condensate energy while $V_{\rm{zero}}\!\left[\langle \bar{\psi}\psi\rangle\right]$ denotes the fermion zero-point energy. The medium potential $V_{\rm{medium}}\!\left[\langle \bar{\psi}\psi\rangle, \ell,\ell^*\right]$ encodes the interactions between the chiral and gauge sector which arises from an integration over the quark fields coupled to a background gauge field. Each part is described in detail in the following sections.

Before moving on it is important to note that PNJL models are non-renormalizable due to the multi-fermion interactions. We truncate the models to the six-fermion operators at most and use a 3D cutoff to obtain meaningful predictions from divergent momentum integrals. The cutoff $\Lambda$ should be understood as a model parameter~\cite{Klevansky:1992qe}. This regularization scheme is most convenient for finite-temperature computation and has been widely used in the NJL literature.

\subsection{Polyakov-Loop Potential}
Pisarski proposed the Polyakov-Loop Model (PLM) in \cite{Pisarski:2000eq, Pisarski:2001pe} as an effective field theory to describe the confinement-deconfinement phase transition of $SU(N)$ gauge theories. The fundamental (traced) Polyakov loop $\ell$ plays the role of the order parameter ($\text{Tr}_c$ denotes the trace in colour space)
\begin{align}
	{\ell}\left(\vec{x}\right)=\frac{1}{N_c}{\rm Tr}_c[{\bf L}]\,,
	\label{eq:Polyakov_Loop}
\end{align}
where
\begin{align}
	{\bf L}(\vec{x})={\mathcal P}\exp\!\left[i\int_{0}^{1/T} \!\! A_{4}(\vec{x},\tau)\,\mathrm d\tau\right],
	\label{eq:thermalWilson}
\end{align}
is the thermal Wilson line at temperature $T$, $\cal P$ denotes the path ordering along the time direction, and $A_4$ is the Euclidean temporal component of the gauge field in the fundamental representation (with the gauge coupling absorbed)\footnote{Note that for all three models, 3F3, 3G1 and 3S1, the Polyakov-loop potential is a potential for the Polyakov loop in the fundamental representation.}. The symbols $\vec{x}$ and $\tau$ denote a spatial point and the Euclidean time, respectively. Note that an ordinary gauge transformation of the gauge field reads
\begin{align}
	A_\mu(x)\rightarrow A'_\mu(x)=V(x)\left[A_\mu(x)+i\partial_\mu\right]V^\dagger(x)\,,
	\label{eq:gt}
\end{align}
with the transformation matrix $V(x)\in SU(N_c)$. A centre symmetry transformation is defined to be a transformation in the form of \cref{eq:gt} with $V(x)$ satisfying a twisted boundary condition ($\beta\equiv 1/T$)~\cite{Fukushima:2017csk, Greensite:2011zz}
\begin{align}
	V(x_4=\beta)&=z_k\cdot V(x_4=0) \,,
	&
	z_k&=e^{i 2\pi k/N_c}\,,
\end{align}
for $k=0,1,...,N_c-1$. Such a centre symmetry transformation preserves the periodic boundary condition for $A_\mu$ along the time direction. The fundamental Polyakov loop transforms non-trivially under the $Z_{N_c}$ centre symmetry
\begin{align}
	\ell&\rightarrow z_k\ell\,,
	&
	k&=0,1,\ldots,N_c-1\,.
	\label{eq:Zn_transformation}
\end{align}
The connection between the centre symmetry and confinement is due to the fact that the free energy of a single static quark in the fundamental representation $F_q$ is related to the thermal average of the traced fundamental Polyakov loop~\cite{Fukushima:2017csk, Greensite:2011zz}
\begin{align}
	\exp(-\beta F_q)\propto \langle \ell\rangle\,.
\end{align}
So an unbroken centre symmetry implies the vanishing of $\langle \ell\rangle$, which in turn implies $F_q$ is infinite, and vice versa. Note the above discussion of centre symmetry applies to the pure-glue sector. The coupling to dynamical fermions may or may not explicitly break the centre symmetry, depending on the quark representation~\cite{Greensite:2011zz}. For the three models in \cref{tab:3models}, fermions in the fundamental and two-index symmetric representation of the $SU(3)$ gauge group leads to explicit breaking of the centre symmetry, while fermions in the adjoint representation preserve the centre symmetry. This is indicated in the ``centre symmetry" column of \cref{tab:3models}.

We adopt the Polyakov gauge~\cite{Weiss:1980rj, Fukushima:2003fw} in which $A_4$ is diagonal and static. Also, in the spirit of mean-field approximation, we consider $A_4$ to be spatially homogeneous~\cite{Meisinger:2001cq}. Then $\ell$ is independent of $\vec{x}$. Introducing also the conjugate traced Polyakov loop $\ell^*$ and the notation $|\ell|^2\equiv\ell\ell^*$, the simplest effective potential preserving the $Z_{N_c}$ symmetry in the polynomial form is given by
\begin{align}
	V_{\rm{PLM}}^{\mathrm{(poly)}}=T^4\left(-\frac{b_2(T)}{2}|\ell|^2+b_4|\ell|^4-b_3\!\left(\ell^{N_c}+\ell^{*N_c}\right)\right),
	\label{eq:PLM_potential}
\end{align}
where
\begin{align}
	b_2(T)=a_0+a_1\!\left(\frac{T_0}{T}\right)\!+a_2\!\left(\frac{T_0}{T}\right)^{\!2}\!
	+a_3\!\left(\frac{T_0}{T}\right)^{\!3}\!+a_4\!\left(\frac{T_0}{T}\right)^{\!4}\!\!.
	\label{eq:b2}
\end{align}
We have chosen the coefficients $b_3$ and $b_4$  to be temperature independent following the treatment in \cite{Ratti:2005jh, Fukushima:2017csk, Huang:2020crf}, and also neglected higher orders in $\lvert\ell\rvert$.

\begin{table}
	\begin{tabular}{|x{2.3cm}|x{.75cm}|x{.75cm}|x{.75cm}|x{.75cm}|x{.75cm}|x{.75cm}|x{.75cm}|}
		\hline
		Parametrization & $a_0$ & $a_1$ & $a_2$ & $a_3$ & $a_4$ & $b_3$ & $b_4$ \\
		\hline
		polynomial & 3.72 & -5.73 & 8.49 & -9.29 & 0.27 & 2.40 & 4.53 \\
		\hline
		logarithmic & 4.26 & -6.53 & 22.8 & -4.10 & & -1.77 & \\
		\hline
	\end{tabular}
	\caption{Parameters in the effective Polyakov-loop potentials, see \cref{eq:PLM_SU3_Log,eq:PLM_potential}.}
	\label{tab:best-fit}
\end{table}

For the $SU(3)$ case, there is also an alternative logarithmic parameterization which includes the information on the Haar measure\footnote{The Haar measure in the context of confinement physics is the Jacobian of the variable transformation from the gauge potential to the Polyakov loop~\cite{Reinhardt:1996fs, Fukushima:2003fm}.}, see e.g.~\cite{Fukushima:2017csk, Roessner:2006xn}, given by
\begin{align}
	\label{eq:PLM_SU3_Log}
	V_{\rm{PLM}}^{(3\mathrm{log})}&=T^4\bigg(\!-\frac{a(T)}{2}|\ell|^2 \\
	&\quad +b(T)\ln\!\left[1-6|\ell|^2+4(\ell^{*3}+\ell^{3})-3|\ell|^4 \right]\!\bigg),\notag
\end{align}
with
\begin{align}
	\label{eq:ab-log}
	a(T) & =a_0+a_1\!\left(\frac{T_0}{T}\right)+a_2\!\left(\frac{T_0}{T}\right)^{\!2}+a_3\!\left(\frac{T_0}{T}\right)^{\!3}\!, \notag \\
	b(T) & = b_3\!\left(\frac{T_0}{T}\right)^{\!3}\!.
\end{align}
The $a_i,\,b_i$ coefficients in  $V_{\text{PLM}}^{\text{(poly)}}$ and $V_{\text{PLM}}^{(3\text{log})}$ have been determined with a dedicated fit in \cite{Huang:2020crf} to available pure-glue lattice data \cite{Panero:2009tv}. The results are shown in \cref{tab:best-fit}

The temperature $T_0$ in \cref{eq:b2,eq:ab-log} is identical to the critical temperature of the confinement phase transition $T_c$, in the pure-glue case. In the presence of fermions, the relation of $T_0$ to $T_c$ is slightly modified, which we will discuss later.

\subsection{Condensate Energy}
\label{subsec:2D}
The condensate energy $V_\text{cond}\!\left[\langle \bar{\psi}\psi\rangle\right]$ can be viewed as a tree-level contribution from the chiral condensate $\langle\bar{\psi}\psi\rangle$ to the grand potential $V_{\rm{PNJL}}$.
It can be derived in a self-consistent mean-field approximation~\cite{Kunihiro:1983ej, Hatsuda:1994pi, Helmboldt:2019pan} in which one introduces auxiliary fields for the condensate and splits the original Lagrangian into a mean-field part $\mL_{\rm{MFA}}$ and a residual interacting part $\mL_{\rm{res}}$ so that in the Bogoliubov-Valatin (BV) vacuum defined by a set of self-consistent conditions (SCC), $\mL_{\rm{res}}$ has vanishing expectation value, and the SCC coincide with the equations of motion for the auxiliary fields derived from $\mL_{\rm{MFA}}$. For computing the condensate energy, the procedure can be simplified to
a linearization of the fermion bilinears around the condensate~\cite{Asakawa:1989bq, Buballa:2003qv}. Moreover, as explained in \cref{subsec:2B}, we only need to consider direct terms.

Here we outline the derivation of the condensate energy for the 3F3 case. The only relevant condensates are
$\langle\bar{u}u\rangle$, $\langle\bar{d}d\rangle$, and $\langle\bar{s}s\rangle$. In $\mL_{\rm{4F}}$, the condensate energy then comes from the combination
\begin{align}
	&(\bar{\psi}\lambda^0\psi)^2+(\bar{\psi}\lambda^3\psi)^2+(\bar{\psi}\lambda^8\psi)^2 \notag \\
	&\qquad=2(\bar{u}u)^2+2(\bar{d}d)^2+2(\bar{s}s)^2\,.
\end{align}
Then the trick is to rewrite $(\bar{u}u)^2$ as
\begin{align}
	\left(\bar{u}u\right)^2&=\left[\left(\bar{u}u-\langle\bar{u}u\rangle\right)+\langle\bar{u}u\rangle\right]^2 \notag \\
	&=\left(\bar{u}u-\langle\bar{u}u\rangle\right)^2+2\langle\bar{u}u\rangle\left(\bar{u}u-\langle\bar{u}u\rangle\right)
	+\langle\bar{u}u\rangle^2 \notag \\
	&\simeq -\langle\bar{u}u\rangle^2+2\langle\bar{u}u\rangle\bar{u}u\,,\label{eq:matrick}
\end{align}
where in the last step the $\left(\bar{u}u-\langle\bar{u}u\rangle\right)^2$ term is dropped in the spirit of the mean-field approximation. In the remaining terms, the $2\langle\bar{u}u\rangle\bar{u}u$ term contributes to the constituent quark mass of $u$ which plays an important role in determining the zero-point energy and medium part of
the potential. The $-\langle\bar{u}u\rangle^2$ term leads to a contribution to the condensate energy. Similar procedures can be applied to $(\bar{d}d)^2$ and $(\bar{s}s)^2$, and to $\mL_{\rm{6F}}$. In the chiral limit, the condensates should exhibit flavour universality, therefore we introduce, in the 3F3 case
\begin{align}
	\mathrm{3F3:}\;\signob\equiv\langle\bar{u}u\rangle=\langle\bar{d}d\rangle=\langle\bar{s}s\rangle
	=\frac{1}{3}\langle\bar{\psi}\psi\rangle\,.
	\label{condensate_3F3}
\end{align}
The condensate energy is then
\begin{align}
	\mathrm{3F3:}\;V_{\mathrm{cond}}=6\Gnob_S\signob^2+\frac{1}{2}\Gnob_D\signob^3\,,
	\label{eq:vcond3F3}
\end{align}
which is consistent with \cite{Fukushima:2008wg, Helmboldt:2019pan} after conversion of notations and conventions. The $\signob^3$ term that originates from the 't Hooft determinental interaction turns out to be an important driving force for a first-order chiral phase transition. In the 3G1 and 3S1 cases, we define
\begin{align}
	\mathrm{3G1\;\;and\;\;3S1:}\,\signob\equiv\langle\bar{\psi}\psi\rangle\,,
	\label{condensate_3G1}
\end{align}
and their condensate energies are found to be
\begin{align}
	\mathrm{3G1:}&\,V_{\mathrm{cond}}=\Gnob_S\signob^2\,, \notag  \\
	\mathrm{3S1:}&\,V_{\mathrm{cond}}=\frac{1}{4}\Gnob_S\signob^2\,.
	\label{eq:vcondGS}
\end{align}
In these cases, the 't Hooft determinantal interaction is associated with some very high-dimensional operator which we neglect in the current approximation, and thus there is no $\signob^3$ term.

\subsection{Zero-Point Energy and Medium Potential}
\label{subsec:2E}
In the mean-field approximation outlined in \cref{subsec:2D}, the effects of the multi-fermion interaction terms $\mL_{\rm{4F}}$ and $\mL_{\rm{6F}}$ boil down to a contribution to the condensate energy shown in \cref{eq:vcond3F3,eq:vcondGS}, and a contribution to the constituent quark mass as discussed below \cref{eq:matrick}. With the inclusion of the constituent quark mass, the covariant kinetic term for the quark field shown in \cref{eq:ckt} becomes
\begin{align}
	\mL'_k&=\bar{\psi}(i\gamma_u D^\mu-M)\psi\,,
	&
	D^\mu&=\partial^\mu-iA^\mu \,,
\end{align}
with $A^\mu=\delta_0^\mu A^0$ and the constituent quark mass $M$ is given by
\begin{align}
	\mathrm{3F3:}\,
	&M=-4\Gnob_S\signob-\frac{1}{4}\Gnob_D\signob^2\,, \label{eq:constituent3F3} \\
	\mathrm{3G1:}\,
	&M=-2\Gnob_S\signob\,, \label{eq:constituent3G1} \\
	\mathrm{3S1:}\,
	&M=-\frac{1}{2}\Gnob_S\signob\,, \label{eq:constituent3S1}
\end{align}
and the result for the 3F3 case is again found to be consistent with \cite{Fukushima:2008wg, Helmboldt:2019pan} after conversion of notations and conventions. In the Polyakov gauge and the mean-field approximation, $A^0=-iA_4$ is diagonal and constant, acting as an imaginary chemical potential. The contribution of $\mL'_k$ to the grand potential can be readily evaluated by a functional integration over the fermion at finite temperature and imaginary chemical potential as $\mL'_k$ is quadratic in the fermion field. The calculation can be found in standard textbooks in thermal field theory~\cite{Kapusta:2006pm}. The resulting contribution to the grand potential can be decomposed into a temperature-independent zero-point energy contribution $V_{\rm{zero}}\!\left[\langle \bar{\psi}\psi\rangle\right]$ and a temperature-dependent thermal quark energy (called medium potential) contribution
$V_{\rm{medium}}\!\left[\langle \bar{\psi}\psi\rangle, \ell,\ell^*\right]$. The expression for the zero-point energy is given by~\cite{Buballa:2003qv, Fukushima:2017csk}
\begin{align}
	V_{\rm{zero}}\!\left[\langle \bar{\psi}\psi\rangle\right]=-{\rm{dim(R)}}\,2 N_f\int\!\! \frac{\mathrm d^3p}{\left(2\pi\right)^3}E_p\,,
\end{align}
where
\begin{align}
	E_p=\sqrt{\vec{p}^{\,2}+M^2}\,,
\end{align}
is the energy of a free quark with constituent mass $M$ and three-momentum $\vec{p}$,
$\rm{dim(R)}$ is the dimension of the quark representation $\rm{R}$, and $N_f$ is the number of Dirac quark flavours. The momentum integral is understood to be regularized by a sharp three-momentum cutoff $\Lambda$, which enters the expression for observables and is thus also a parameter of the theory. The integration can be carried analytically and the result is~\cite{Fukushima:2013rx}
\begin{align}
	V_{\text{zero}}\!\left[\langle \bar{\psi}\psi\rangle\right]
	&=-\frac{\mathrm{dim(R)}N_f\Lambda^4}{8\pi^2}\bigg[(2+\xi^2)\sqrt{1+\xi^2} \notag \\
	&\quad\,+\frac{\xi^4}{2}\ln\frac{\sqrt{1+\xi^2}-1}{\sqrt{1+\xi^2}+1}\bigg],
\end{align}
in which $\xi\equiv\frac{M}{\Lambda}$. The spontaneous chiral symmetry breaking at zero temperature is the result of the interplay between the negative contribution from $V_{\rm{zero}}\!\left[\langle \bar{\psi}\psi\rangle\right]$ which favours large values of $M$, and the positive contribution from $V_{\rm{cond}}\!\left[\langle \bar{\psi}\psi\rangle\right]$ which favours small values of $M$~\cite{Buballa:2003qv}.

The medium potential $V_{\rm{medium}}\!\left[\langle \bar{\psi}\psi\rangle, \ell,\ell^*\right]$ is evaluated to be~\cite{Fukushima:2017csk}
\begin{align}
	\label{eq:medium-pot}
	V_{\rm{medium}}\!\left[\langle \bar{\psi}\psi\rangle, \ell,\ell^*\right]
	=-2N_f T\int\!\! \frac{\mathrm d^3p}{\left(2\pi\right)^3}(S_R+S_R^\dagger)\,,
\end{align}
in which $S_R$ and $S_R^\dagger$ are defined as
\begin{align}
	S_R &\equiv\Tr_C\ln\!\left[1+L_R\exp\!\Big(-\frac{E_p-\mu}{T}\Big)\right], \notag \\
	S_R^\dagger &\equiv\Tr_C\ln\!\left[1+L_R^\dagger\exp\!\Big(-\frac{E_p+\mu}{T}\Big)\right].
\end{align}
Here $\mu$ is the chemical potential that we take to be zero. $L_R$ is the Polyakov loop matrix in the quark representation $\mathrm{R}$, defined in a similar manner to \cref{eq:thermalWilson}
\begin{align}
	L_R(\vec{x})={\mathcal P}\exp\!\left[i\int_{0}^{1/T} \!\! A_{4}^{\mathrm{(R)}}(\vec{x},\tau)\,\mathrm d\tau\right],
\end{align}
with the Euclidean temporal gauge field $A_4$ now taken to be in the representation $\mathrm{R}$. Accordingly, we also define the normalized traced Polyakov loop in the representation $\mathrm{R}$
\begin{align}
	{\ell_R}\left(\vec{x}\right)&=\frac{\Tr_C[L_R]}{\mathrm{dim(R)}}\,,
	&
	{\ell_R^*}\left(\vec{x}\right)&=\frac{\Tr_C[L_R^\dagger]}{\mathrm{dim(R)}}.
\end{align}
We now proceed to evaluate $S_R$ and $S_R^\dagger$ at zero chemical potential. In the spirit of the mean-field approximation, we utilize the properties of the traced Polyakov loops that are satisfied at the saddle point of the grand potential. Especially, we note that the traced Polyakov loop $\ell_R$ at the saddle point is always real\footnote{This is because the saddle-point value of $\ell_R$ is just the thermal average $\langle\ell_R\rangle$ in the presence of zero external source, and thus has the interpretation of the free energy of a static quark in the representation $\mathrm{R}$. Away from the saddle point, $\ell_R$ does not need to be real.}, and becomes equal to its conjugate $\ell_R^*$ at zero chemical potential~\cite{Fukushima:2017csk}. With this in mind, in the grand potential we set $\ell_R=\ell_R^*$~\cite{Kahara:2012yr, Helmboldt:2019pan}, and the $L_R$ in the fundamental representation of $SU(3)$ can be parameterized as
\begin{align}
	L_F=\mathrm{diag}\Big\{e^{i\theta},e^{-i\theta},1\Big\},
	\label{eq:LF}
\end{align}
assuming Polyakov gauge and spatial homogeneity. The phase $\theta$ is then related to $\ell_R$ in the fundamental representation as
\begin{align}
	\cos\theta=\frac{3\ell_F-1}{2}\,,
\end{align}
$S_R$ and $S_R^\dagger$ in the fundamental representation can now be easily evaluated using the parametrization in \cref{eq:LF}, and the result is
\begin{align}
	S_F &=\ln\!\Big[1+(3\ell_F-1)e^{-E_p/T}+e^{-2E_p/T}\Big] \notag \\
	&\quad\,+\ln\!\Big(1+e^{-E_p/T}\Big),
	\label{eq:SF}
\end{align}
with $S_F^\dagger=S_F$. To evaluate $S_R$ and $S_R^\dagger$ in higher representations, we note that $S_R$ is invariant under a similarity transformation in colour $\mathrm{R}$-representation space, thus only the eigenvalues of $L_R$ matter for the calculation. $L_R$ can always be brought into a diagonal form via a similarity transformation, and its diagonal entries become pure phase factors $\exp(i\lambda_j^R),\,j=1,2,...,\mathrm{dim(R)}$ since $L_R$ is unitary. The eigenvalue phases $\lambda_j^R$, with $j=1,2,...,\mathrm{dim(R)}$, are \emph{weights} of the irreducible representation $\mathrm{R}$, and they can be obtained from the weights (i.e.\ eigenvalue phases) of the fundamental representation which we parameterize. For example, in the adjoint representation case, $L_\text{Adj}$ can be parameterized as
\begin{align}
	L_\text{Adj}=\mathrm{diag}\Big\{e^{2i\theta},e^{-2i\theta},e^{i\theta},e^{-i\theta},e^{i\theta},e^{-i\theta},1,1\Big\},
\end{align}
with $\theta$ defined in the parameterization of $L_F$ in \cref{eq:LF}. $S_R$ in the adjoint representation is then computed to be
\begin{align}
	S_\text{Adj} &=2\ln\!\Big(1+e^{-E_p/T}\Big) \notag \\
	&\quad\,+\ln\!\Big[1+(9\ell_F^2-6\ell_F-1)e^{-E_p/T}+e^{-2E_p/T}\Big] \notag \\
	&\quad\,+2\ln\!\Big[1+(3\ell_F-1)e^{-E_p/T}+e^{-2E_p/T}\Big],
	\label{eq:SAdj}
\end{align}
with $S_\text{Adj}^\dagger=S_\text{Adj}$. For the two-index symmetric representation, $L_{S_2}$ can be parameterized as
\begin{align}
	L_{S_2}=\mathrm{diag}\Big\{e^{2i\theta},e^{-2i\theta},1,1,e^{i\theta},e^{-i\theta}\Big\},
\end{align}
with the same $\theta$ introduced in \cref{eq:LF}. $S_R$ in the two-index symmetric representation is then computed to be
\begin{align}
	S_{S_2} &=2\ln\!\Big(1+e^{-E_p/T}\Big) \notag \\
	&\quad\,+\ln\!\Big[1+(9\ell_F^2-6\ell_F-1)e^{-E_p/T}+e^{-2E_p/T}\Big] \notag \\
	&\quad\,+\ln\!\Big[1+(3\ell_F-1)e^{-E_p/T}+e^{-2E_p/T}\Big],
	\label{eq:SS2}
\end{align}
with $S_{S_2}^\dagger=S_{S_2}$. The expressions we obtained for $S_R$ in \cref{eq:SF,eq:SAdj,eq:SS2} agree with \cite{Kahara:2012yr}.

If we consider a gauge group of larger rank, then the fundamental traced Polyakov loop $\ell_F$ alone is not sufficient to describe the medium potential. For example, in the $SU(4)$ case, we need two angles to parameterize $L_F$
\begin{align}
	L_F^{(4)}=\mathrm{diag}\Big\{e^{i\theta_1},e^{-i\theta_1},e^{i\theta_2},e^{-i\theta_2}\Big\},
\end{align}
where the superscript ``$(4)$" is used to indicate the quantities associated with the $SU(4)$ gauge group. The $\theta_1,\theta_2$ angles can be traded for two independent traced Polyakov loops, in the fundamental and two-index antisymmetric representations, respectively
\begin{align}
	\ell_F^{(4)} &=\frac{1}{2}(\cos\theta_1+\cos\theta_2)\,, \notag \\
	\ell_{A_2}^{(4)} &=\frac{2}{3}\cos\theta_1\cos\theta_2 \,.
\end{align}
Then $S_R$ depends on $\ell_F^{(4)}$ and $\ell_{A_2}^{(4)}$ simultaneously. This is true even for fermions in the fundamental representation
\begin{align}
	S_F^{(4)}&=\ln\!\Big[1+4\ell_F^{(4)}(e^{-E_p/T}+e^{-3E_p/T}) \notag \\
	&\quad\,+6\ell_{A_2}^{(4)}e^{-2E_p/T}+e^{-4E_p/T}\Big],
\end{align}
and similarly for $S_R$ in higher representations (with more complicated expressions). This suggests treating the confinement dynamics and the interaction between the quark and gluon sectors not in terms of a single traced Polyakov loop in the fundamental representation but in terms of eigenvalues of the Polyakov-loop matrix, which goes in the line of the matrix-model approach~\cite{Meisinger:2001cq, Dumitru:2010mj, Dumitru:2012fw, Halverson:2020xpg}. Moreover, for the convenience of studying the bubble nucleation, some method needs to be introduced to reduce the multi-variable problem to the tunnelling in a single dimension~\cite{Halverson:2020xpg}. The extension of the current work to these cases will be left for future study.

The medium potential $V_{\rm{medium}}\!\left[\langle \bar{\psi}\psi\rangle, \ell,\ell^*\right]$ does not contain ultraviolet (UV) divergence and there are different procedures on the market regarding the regularization of this contribution, even within a 3D momentum cutoff framework. For example, in \cite{Fukushima:2003fw}, no momentum cutoff is imposed on the medium potential and the 3D momentum is integrated to infinity. On the other hand, in \cite{Hansen:2006ee}, a sharp 3D momentum cutoff has been employed everywhere, including the medium potential. The choice is motivated by the authors' wish to describe certain mesonic properties. When it comes to quarks in higher representations, \cite{Kahara:2012yr} regulates the medium potential by introducing a momentum-dependent four-fermion coupling
\begin{align}
	\label{eq:mom-dep-cutoff}
	\Gnob_S(|\vec{p}\,|)=\Gnob_S\,\theta(\Lambda-|\vec{p}\,|)\,,
\end{align}
which implies that for three-momentum larger than $\Lambda$, the medium potential is not set to zero, but rather computed as if the quarks have zero constituent mass. It is found in \cite{Kahara:2012yr} that such a regularization treatment is needed to obtain clearly separated confinement and chiral phase transition in the case of adjoint quarks\footnote{Interestingly, we observe that without this regularization treatment, it is possible to obtain a first-order chiral phase transition for sufficient large $G_S$ while the confinement and chiral phase transition become strongly correlated like in the fundamental case.}.

\subsection{Model Parameters and Observables}
Apart from the coefficients in the Polyakov-loop potential, the PNJL models we are considering have only two parameters ($\Gnob_S$ and $\Lambda$) in the 3G1 and 3S1 cases, and three parameters ($\Gnob_S$, $\Gnob_D$, and $\Lambda$) in the 3F3 case. In principle, these parameters should be determined from observables (meson masses, decay constants) measured from experiments or predicted in lattice calculations. However, because we work in the chiral limit, even in the 3F3 case it is difficult to determine the parameters precisely. In \cite{Helmboldt:2019pan}, four benchmark points were chosen to study the 3F3 model in the chiral limit. We deem it reasonable in the 3F3 case to use the parameters corresponding to physical real-world values as a reference and then investigate variations away from the physical point by some amount. The values for such a choice can be found in \cite{Buballa:2003qv}, see also \cite{Abuki:2010jq}. In the 3G1 and 3S1 cases, however, there are no clear guidelines to determine the parameter and we thus allow the parameters to vary in a larger range.

Nonetheless, we provide formulae for a set of observables to gain more physical insight from the model parameters that we use, and also to facilitate future comparison of computations done in different approaches (because a meaningful comparison should be carried out at the same value of observables). The set of observables that we consider include the chiral condensate $\signob$, the constituent quark mass $M$, the pion-decay constant $f_\pi$, and the $\sigma$-meson mass $m_\sigma$.

The chiral condensate $\signob$ is determined from the saddle point equation at zero temperature and chemical potential
\begin{align}
	\frac{\partial(V_{\mathrm{cond}}+V_{\mathrm{zero}})}{\partial\signob}=0\,,
\end{align}
which is just the gap equation of the corresponding NJL model. The constituent quark mass $M$ is then given by \cref{eq:constituent3F3,eq:constituent3G1,eq:constituent3S1}.

The pion-decay constant is determined from the vacuum to one-pion axial-vector matrix element, and we use the normalization in \cite{Klevansky:1992qe}. In the 3D momentum cutoff scheme, it is given by
\begin{align}
	f_\pi =M\sqrt{\frac{\mathrm{dim(R)}}{2\pi^2}} \sqrt{\mathrm{arcsinh}\frac{\Lambda}{M}-\frac{\Lambda}{\sqrt{\Lambda^2+M^2}}}\,.
\end{align}
Finally, the $\sigma$-meson mass $m_\sigma$ is the root of the 1PI $\sigma$-$\sigma$ 2-point function $\Gamma_{\sigma\sigma}$ whose expressions are lengthy and thus given in \cref{appendix:wfr}. It turns out that in the 3G1 and 3S1 cases the $\sigma$-meson mass is simply twice the constituent quark mass.

\subsection{Comments on Universality}
\label{subsec:2G}
We have collected various pieces of the PNJL grand potential for the three models of our interest. Before moving to the exploration of the phase transitions based on the expressions for the grand potential, let us comment on the relation between the universality argument and our analysis. There is a nice summary of the logic of the universality argument in \cite{Basile:2005hw}:
\begin{enumerate}
	\item One first assumes the phase transition is continuous. The asymptotic critical
	behaviour must be associated with a 3D universality class with the same symmetry breaking
	pattern as the original theory.
	\item The existence of such a 3D universality class can be investigated by considering the
	most general Landau-Ginzburg-Wilson $\Phi^4$ theory compatible with the given symmetry breaking pattern.
	\item The critical behaviour at a continuous transition is determined by the fixed points
	(FPs) of the renormalization group (RG) flow: the absence of a stable FP generally implies first-order transitions. However, if a stable FP exists, the phase transition can be of second-order or first-order (if the theory is outside the attractive domain of the FP).
\end{enumerate}
RG predictions for the type of chiral phase transitions in $SU(3)$ QCD theories with quarks in the complex or real representations have also been summarized in \cite{Basile:2005hw}. The universality analysis can also be carried out for the confinement phase transition, see \cite{Fukushima:2013rx} for a summary. Among the three models we are considering, 3F3 is predicted to exhibit a first-order chiral phase transition. On the other hand, for 3G1 and 3S1, no definite prediction can be made. It is therefore well-motivated to explore what type of phase transitions these models exhibit using an effective theory approach like PNJL.

For quarks in the fundamental representation, the universality argument predicts a first-order phase transition also for a number of Dirac flavours larger than $3$ (but below the lower boundary of the conformal window). On the other hand, our study suggests that for $N_f=4$ the PNJL approach does not seem to exhibit a first-order chiral phase transition even if the model parameters are allowed to vary in a large range. This does not mean the universality argument is invalid. Rather, it is likely that this points to the possibility that the PNJL approach fails to model the phase-transition dynamics faithfully in such a case. This motivates further modelling of the phase transition using alternative effective theories in order to deliver a first-order chiral phase transition compatible with the prediction of universality.

\section{Confinement and Chiral Phase Transitions}
\label{sec:phase-transition}
In this section, we discuss the nature of the confinement and chiral phase transition for the models studied in this work. We start with some cosmological considerations and the discussion of order parameters, followed by a generic review of the bubble nucleation and the resulting GW spectrum.

\subsection{Cosmological considerations}
\label{sec:cosmo}
We briefly discuss here the cosmological constraints on the dark pions in our model. This can be divided into the two cases of massless and massive dark pions.

\textbf{Massless dark pions.} The dark pions play the role of dark radiation which is strongly constrained by the cosmic microwave background (CMB). Following the nice discussion in \cite{Bai:2018dxf}, there are two viable cases:
\begin{enumerate}
	\item The dark sector and visible sector are thermalized in the very early universe but decouple prior to the electroweak scale $T_\text{ew}$. In addition, the chiral phase transition should happen even before the decoupling. In this case, for the $SU(3)$ gauge group with fermions in the fundamental and two-index symmetric representation, only $1\leq N_f\leq3$ Dirac flavours are viable while in the adjoint representation, only one Dirac flavour is viable. Thus our choice of $N_f=3$ and $N_f=1$ in respectively fundamental and adjoint (two-index symmetric) representations is valid.
	\item The dark sector and visible sector never thermalize. In this case, the key parameter from the CMB constraint i.e.~the ratio between the hidden and visible sector temperature during the CMB epoch $T_d\left(t_{\rm{CMB}}\right)/T_{v}\left(t_{\rm{CMB}}\right)$ can be made arbitrarily small and thus avoid the constraints from CMB. Note that the GW signal is suppressed if the dark temperature is much colder than the visible temperature \cite{Breitbach:2018ddu, Garani:2021zrr} and thus we do not consider this case.
\end{enumerate}

\textbf{Massive dark pions.} This scenario is less constrained and can be achieved by adding small explicit quark masses. We assume the dark gauge sector is in thermal equilibrium with the SM at early times before the Big-Bang Nucleosynthesis (BBN) and that the pseudo-Nambu-Goldstone bosons (pNGBs) have decay channels into lighter SM particles. As long as they have a mass larger than a few $\MeV$s and can decay before BBN, these pNGBs do not cause conflicts with cosmological and laboratory constraints~\cite{Kawasaki:1992kg}, which we explain in more detail in \cref{appendix:constraints}.

In this work, we consider the simplest case where at the phase transition the dark and visible sectors are in thermal equilibrium, $T_d=T_v$. This implies that the dark pions can be massless for $T_c > T_\text{ew}$, while they have to be massive for $T_c < T_\text{ew}$.

\begin{figure}[t]
	\includegraphics[width=\linewidth]{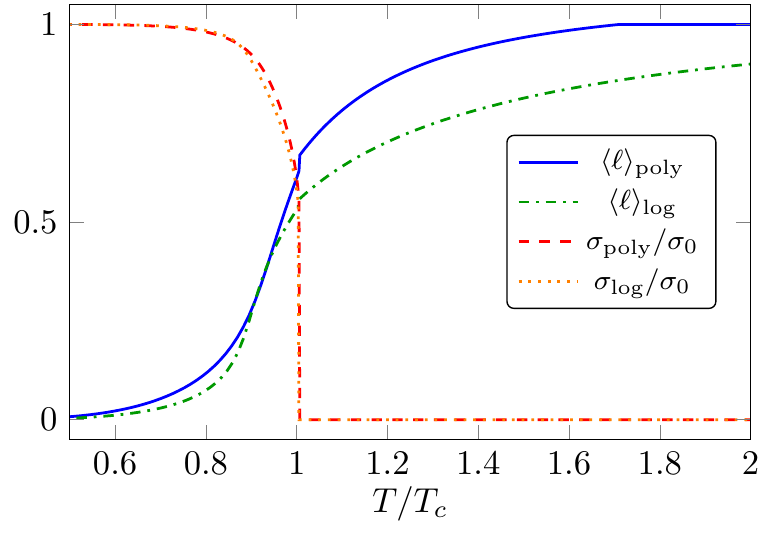}
	\caption{Fundamental representation: expectation value of the Polyakov loop $\langle\ell\rangle$ and the chiral condensate $\sigma$ as a function of temperature. The latter is normalised to its value at vanishing temperature $\sigma_0$. We present the polynomial and logarithmic fitting of the Polyakov-loop potential.
	}
	\label{fig:fund-order}
\end{figure}

\subsection{Order parameters}
\label{sec:order-param}
In this section, we focus on the interplay between chiral and confinement phase transitions for the models studied here,  see \cref{tab:3models}. The order parameter of the confinement phase transition is the Polyakov-loop expectation value, while the chiral condensate is the order parameter for the chiral phase transition. Fermions in the fundamental and two-index symmetric representation explicitly break the $Z_3$ centre symmetry of $SU(3)$ and thus the Polyakov loop is no longer a rigorous order parameter for the confinement phase transition. Nevertheless, the Polyakov loop can still serve as an indicator of a crossover between confinement and deconfinement \cite{Fukushima:2002bk, Ratti:2005jh}.

\begin{figure}[t]
	\includegraphics[width=\linewidth]{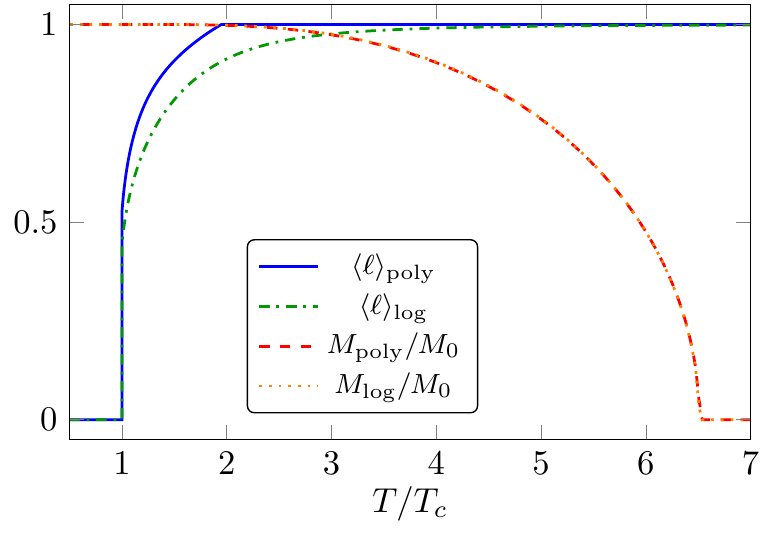}
	\caption{Adjoint representation: expectation value of the Polyakov loop $\langle\ell\rangle$ and the constituent quark mass $M$ as a function of temperature. The latter is normalised to its value at vanishing temperature $M_0$. We present the polynomial  and logarithmic fitting of the Polyakov-loop potential.
	}
	\label{fig:adjoint-order}
\end{figure}

\begin{figure}[t]
	\includegraphics[width=\linewidth]{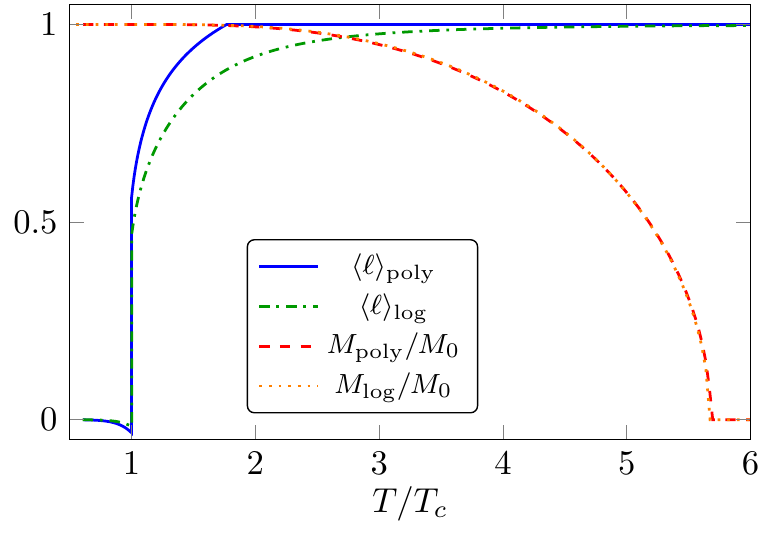}
	\caption{Two-index symmetric representation: expectation value of the Polyakov loop $\langle\ell\rangle$ and the constituent quark mass $M$ as a function of temperature. The latter is normalised to its value at vanishing temperature $M_0$. We present the polynomial and logarithmic fitting of the Polyakov-loop potential.
	}
	\label{fig:sym-order}
\end{figure}

We display the expectation values of the Polyakov loop $\langle\ell\rangle$ and the chiral condensate $\sigma$ (or equivalently of the constituent quark mass $M$) as a function temperature for each case, see \cref{fig:fund-order,fig:adjoint-order,fig:sym-order}. The chiral condensate and the constituent mass are normalised to their respective values at vanishing temperature. We display the polynomial and logarithmic fitting of the effective Polyakov-loop potential, \cref{eq:PLM_SU3_Log,eq:PLM_potential}. In the pure-glue case, both potentials have the property that $\langle \ell\rangle \to 1$ for $T\to\infty$, however, with the addition of the medium potential \cref{eq:medium-pot} this property is lost in the polynomial case. In principle, one would need to refit the coefficients to the lattice data with the refined constraint of $\langle \ell\rangle \to 1$ for $T\to\infty$ that includes the properties in the medium potential. We refrain from doing so since this effect only becomes relevant at large $T$ and is not relevant for the dynamics of the phase transition. For the purpose of the figures, we implement the constraint $\langle \ell \rangle \leq 1$ by hand. We also emphasise that the logarithmic potential naturally implements the constraint $\langle \ell\rangle \to 1$ for $T\to\infty$ and therefore might be better suited for the considerations in this work.

\textbf{Fundamental representation $SU(3)$ with $N_f=3$ (\cref{fig:fund-order}):}
The chiral condensate has a discontinuity at the critical temperature and thus we have a first-order chiral phase transition. The Polyakov loop expectation value undergoes a cross over (there is a small discontinuity at $T_c$ due to the discontinuity in $\sigma$). The confinement crossover happens roughly at the same temperature as the first-order chiral phase transition.

\textbf{Adjoint Representation $SU(3)$ with $N_f=1$ (\cref{fig:adjoint-order}):}
The fermions in the adjoint representation do not break the $Z_3$ centre symmetry and thus the Polyakov loop expectation value remains a good order parameter for the confinement phase transition. We find a first-order confinement phase transition, while the chiral phase transition is of second order and happens at much larger temperatures.

\textbf{Two-index symmetric Representation:~$SU(3)$ with $N_f=1$ (\cref{fig:sym-order}):}
The two-index symmetric representation is a very interesting case. In principle, the centre symmetry is explicitly broken by the fermions and thus there should be no confinement phase transition. However, it turns out that the centre symmetry is only weakly broken \cite{Kahara:2012yr}. The amount of symmetry breaking is characterised by $1/M$ where $M$ is the constituent quark mass. The latter is rather large in the two-index symmetric representation, see \cref{tab:GW-param-obs}, and consequently, the centre symmetry is only softly broken. As in the adjoint case, the chiral phase transition is of second-order and happens at larger temperatures. Therefore the centre symmetry is almost restored at $T_c$, and we observe a first-order confinement phase transition. The small negative dip of the Polyakov loop expectation value is precisely due to the breaking of the centre symmetry induced via the medium potential \cref{eq:medium-pot}.

\subsection{Bubble Nucleation}
\label{sec:bubble-nucl}
In case of a first-order phase transition, the transition occurs via bubble nucleation and it is essential for the understanding of the dynamics to compute the nucleation rate. The tunnelling rate due to thermal fluctuations per unit volume as a function of the temperature from the metastable vacuum to the stable one is suppressed by the three-dimensional Euclidean action $S_3(T)$ \cite{Coleman:1977py, Callan:1977pt, Linde:1980tt, Linde:1981zj}
\begin{align}
	\Gamma(T)=T^4\left(\frac{S_3(T)}{2\pi T}\right)^{\!3/2} e^{-S_3(T)/T}.
	\label{eq:decay_rate}
\end{align}
The three-dimensional Euclidean action reads
\begin{align}
	S_3(T)=4\pi\!\int_0^\infty \!\!\mathrm dr\,r^2\!\left[\frac{1}{2}\!\left(\frac{\mathrm d\rho}{\mathrm dr}\right)^{\!2} +V_\text{eff}(\rho,T)\right],
	\label{eq:Euclidean_Action_general}
\end{align}
where $\rho$ denotes a generic scalar field with mass dimension one, $\left[\rho\right]=1$, and $V_\text{eff}$ denotes its effective potential. In our case, the effective potential depends on two scalar fields, the Polyakov loop $\ell$ and the chiral condensate $\sigma$. Which field takes the leading role depends on whether we have a first-order confinement or chiral phase transition and therefore we discuss them separately.

\textbf{Confinement phase transition:}
The phase transition is described by the Polyakov loop~$\ell$ and it is a first-order phase transition in the adjoint and two-index symmetric case, see \cref{fig:adjoint-order,fig:sym-order}. In both cases, the second-order chiral phase transition is at significantly higher temperatures and has already been completed. Therefore, we can work in the approximation that $\signob$ is constant. Note also that $\ell$ is dimensionless while $\rho$ in \cref{eq:Euclidean_Action_general} has mass dimension one. We therefore rewrite the scalar field as $\rho=\ell\, T$ and convert the radius into a dimensionless quantity $r'=r\, T$. Thus, the action becomes
\begin{align}
	S_3(T)=4\pi T \!\int_0^\infty \!\!\mathrm dr'\, r'^2\!&\left[\frac{1}{2}\left(\frac{\mathrm d\ell}{\mathrm dr'}\right)^{\!2} +V'_\text{eff}(\ell,T)\right]\,,
	\label{eq:Euclidean_Action}
\end{align}
which has the same form as \cref{eq:Euclidean_Action_general}. Here, $V'_\text{eff} (\ell,T) = V_\text{eff}(\ell,T)/T^4$ is dimensionless. The bubble profile (instanton solution) is obtained by solving the equation of motion of the action in \cref{eq:Euclidean_Action}
\begin{align}
	\frac{\mathrm d^2\ell(r')}{\mathrm dr'^2}+\frac{2}{r'}\frac{\mathrm d\ell(r')}{\mathrm dr'}-\frac{\partial V_\text{eff}'(\ell,T)}{\partial\ell}=0\,,
	\label{eq:bounce-solution}
\end{align}
with the associated boundary conditions
\begin{align}
	\frac{\mathrm d\ell(r'=0,T)}{\mathrm dr'}&=0\,,
	&
	\lim_{r'\rightarrow 0} \ell(r',T)&=0\,.
\end{align}
To attain the solutions, we used the method of overshooting/undershooting and employ the \texttt{Python} package \texttt{CosmoTransitions} \cite{Wainwright:2011kj}.

\textbf{Chiral phase transition:}
The chiral phase transition is described by the chiral condensate $\signob$, see \cref{condensate_3F3,condensate_3G1}. In the three models studied here, we only find a first-order chiral phase transition in the fundamental case, see \cref{fig:fund-order}. In order to have a field with mass dimension one, we define
\begin{align}
	\sigbar&\equiv-4\Gnob_S\signob\,.
\end{align}
We work in the mean-field approximation where we evaluate the Polyakov loop $\ell$ for given values of $\bar \sigma$ and $T$ at the minimum of the effective potential. Thus the potential becomes a function of only $\left(\sigbar,T\right)$, $V_\text{eff}(\bar \sigma, T) = V_\text{eff}(\bar \sigma, T, \ell_\text{min}(\bar \sigma, T))$.

\begin{figure}[t]
	\includegraphics[width=\linewidth]{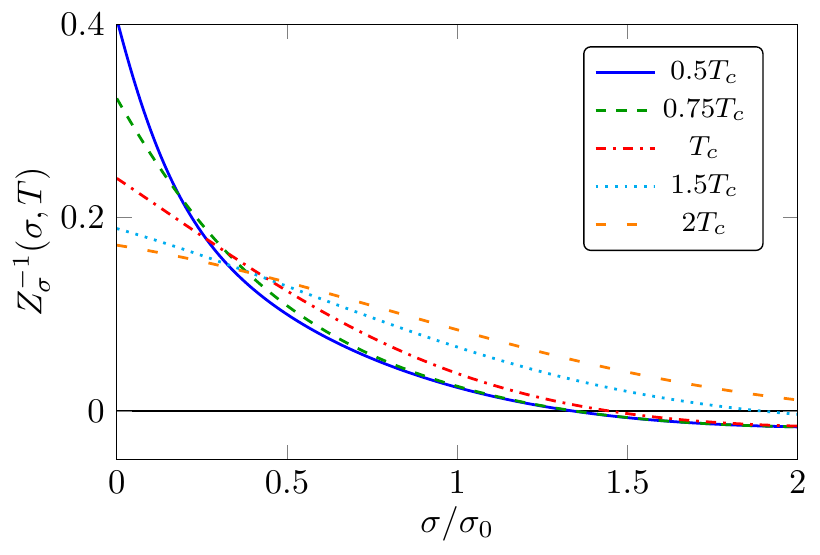}
	\caption{Wave-function renormalization $Z_\sigma^{-1}$ for different temperatures as a function of the chiral condensate $\sigma$ normalised to its value at zero temperature $\sigma_0$.
	}
	\label{fig:Z_factor}
\end{figure}

Since $\sigbar$ is not a fundamental field, we have to include  its wave-function renormalization $Z_\sigma$, see \cref{appendix:wfr} for more details. In \cref{fig:Z_factor}, we display the wave-function renormalization as a function of the chiral condensate and the temperature. The three-dimensional Euclidean action is slightly modified \cite{Helmboldt:2019pan}
\begin{align}
	S_3(T)=4\pi\!\int_0^\infty \!\!\mathrm dr\,r^2\!\left[\frac{Z_{\sigma}^{-1}}{2}\!\left(\frac{\mathrm d\sigbar}{\mathrm dr}\right)^{\!2} +V_\text{eff}(\sigbar,T)\right].
	\label{eq:Euclidean_Action_fund}
\end{align}
The bubble profile is obtained by solving the equation of motion of the action in \cref{eq:Euclidean_Action_fund} and is given by
\begin{equation}
	\frac{\mathrm d^2\sigbar}{\mathrm dr^2}+\frac{2}{r}\frac{\mathrm d\sigbar}{\mathrm dr}-\frac{1}{2}\frac{\partial\log Z_{\sigma}}{\partial\sigbar}\left(\frac{\mathrm d\sigbar}{\mathrm dr}\right)^2=Z_{\sigma}\frac{\partial V_\text{eff}}{\partial\sigbar}\,,
	\label{EOM_fund}
\end{equation}
with the associated boundary conditions
\begin{align}
	\frac{\mathrm d\sigbar(r=0,T)}{\mathrm dr}&=0\,,
	&
	\lim_{r\rightarrow \infty} \sigbar(r,T)&=0\,.
\end{align}
For $Z_{\sigma}=1$, \cref{EOM_fund} simplifies to \cref{eq:bounce-solution}. We use again the overshooting/undershooting method and employ the \texttt{Python} package \texttt{CosmoTransitions} \cite{Wainwright:2011kj} with a modified equation of motion. We substitute the solved bubble profile $\sigbar(r,T)$ into the three-dimensional Euclidean action \cref{eq:Euclidean_Action_fund} and, after integrating over $r$, $S_3$ depends only on $T$.

\subsection{Gravitational-wave parameters}
\label{sec:GW-params}

\subsubsection{Inverse duration time}
\label{sec:beta}
An important parameter for determining the GW spectrum is the rate at which the phase transition completes. For sufficiently fast phase transitions, the decay rate can be approximated by
\begin{align}
	\Gamma(T) \approx \Gamma(t_*) e^{\beta (t-t_*)}\,,
	\label{eq:approx-Gamma}
\end{align}
where $t_*$ is a characteristic time scale for the production of GWs to be specified below. The inverse duration time then follows as
\begin{align}
	\beta= - \frac{\mathrm d}{\mathrm dt} \frac{S_3(T)}{T}\bigg\vert_{t=t_*}\,.
	\label{beta}
\end{align}
The dimensionless version $\tilde \beta$ is defined relative to the Hubble parameter $H_*$ at the characteristic time $t_*$
\begin{align}
	\tilde \beta = \frac{\beta}{H_*}=T\frac{\mathrm d}{\mathrm dT}\frac{S_3(T)}{T}\bigg\vert_{T=T_*}\,,
	\label{eq:beta-tilde}
\end{align}
where we used that $\mathrm dT/\mathrm dt = -H(T)T$. Note that here we assumed that the temperature in the hidden and visible sectors are the same, $T_d=T_v$.

The phase-transition temperature $T_*$ is often identified with the nucleation temperature $T_n$, which is defined as the temperature at which the rate of bubble nucleation per Hubble volume and time is approximately one, i.e.\ $\Gamma/H^4\sim \mathcal{O}(1)$. More accurately one can use the percolation temperature $T_p$, which is defined as the temperature at which the probability to have the false vacuum is about $0.7$. For very fast phase transitions, as in our case, the nucleation and percolation temperature are almost identical $T_p\lesssim T_n$. However, even a small change in the temperature leads to an exponential change in the vacuum decay rate
$\Gamma$, see \cref{eq:approx-Gamma}, and consequently, we use the percolation temperature throughout this work. We write the false-vacuum probability as~\cite{Guth:1979bh, Guth:1981uk}
\begin{align}
	P(T) = e^{-I(T)}\,,
\end{align}
with the weight function \cite{Ellis:2018mja}
\begin{align}
	I(T)=\frac{4\pi}{3} \int^{T_c}_T \! \!\mathrm dT'\frac{\Gamma(T')}{H(T')T'{}^4} \left( \int^{T'}_{T}\!\!\mathrm dT''\frac{v_w(T'')}{H(T'')}\right)^{\!3}.
	\label{eq:Tp}
\end{align}
The percolation temperature is defined by $I(T_p)=0.34$, corresponding to $P(T_p)= 0.7$ \cite{Rintoul_1997}. Using $T_*=T_p$ in \cref{eq:beta-tilde} yields the dimensionless inverse duration time. We will see that all phase transitions considered here have very fast rates, $\tilde \beta \sim \mathcal{O}(10^4)$.

\subsubsection{Energy budget}
We define the strength parameter $\alpha$ from the trace of the energy-momentum tensor $\theta$ weighted by the enthalpy
\begin{align}
	\alpha=\frac{1}{3}\frac{\Delta\theta}{w_+}=\frac{1}{3}\frac{\Delta e\,-\,3\Delta p}{w_+}\,,
	\label{alpha_def}
\end{align}
where $\Delta X= X^{(+)}-X^{(-)}$ for $X = (\theta$, $e$, $p$) and $(+)$ denotes the meta-stable phase (outside of the bubble) while $(-)$ denotes the stable phase (inside of the bubble). The relations between enthalpy $w$, pressure $p$, and energy $e$ are given by
\begin{align}
	w&=\frac{\partial p}{\partial \ln T}\,,
	&
	e&=\frac{\partial p}{\partial \ln T} -p\,.
	\label{eq:enthalpy}
\end{align}
These are hydrodynamic quantities and we work in the approximation where do not solve the hydrodynamic equations but instead extract them from the effective potential with
\begin{align}
	p^{(\pm)}&=-V_{\text{eff}}^{(\pm)}\,.
	\label{eq:def-p}
\end{align}
This treatment should work well for the phase transitions considered here, see \cite{Giese:2020rtr, Giese:2020znk, Wang:2020nzm}. With \cref{eq:enthalpy,eq:def-p}, $\alpha$ is given by
\begin{align}
	\alpha=\frac{1}{3}\frac{4\Delta V_\text{eff}-T\frac{\partial \Delta V_\text{eff}}{\partial T}}{-T\frac{\partial V_{\text{eff}}^{(+)}}{\partial T}}\,.
	\label{eq:alpha}
\end{align}
In case of the confinement phase transition, we find that the contribution from $\Delta V_\text{eff}$ is negligible since $e_+\gg p_+$ and therefore $\alpha \approx 1/3$. In the case of the chiral phase transition, we find smaller values, $\alpha \sim \mathcal{O}(10^{-2})$, which relates to the fact that there are more relativistic d.o.f.s participating in the phase transition. Note that relativistic SM d.o.f.s do not contribute to our definition of $\alpha$ since they are fully decoupled from the phase transition. The dilution due to the SM d.o.f.s is included at a later stage, see \cref{sec:spectrum}.

\subsubsection{Bubble-wall velocity}
\label{sec:wall-velocity}
We treat the bubble-wall velocity $v_w$ as a free parameter. A reliable estimate of the wall velocity would require a detailed analysis of the pressure and friction on the bubble wall. The latter is typically evaluated in an expansion of $1 \to n$ processes \cite{Bodeker:2009qy, Bodeker:2017cim, Cai:2020djd, Baldes:2020kam, Azatov:2020ufh, Wang:2020zlf}. In our case, we have a strongly coupled system and most likely a fully non-perturbative analysis would be necessary to determine the friction.

We make the assumption that the wall velocity is larger than the speed of sound, $v_w \geq c_s=1/\sqrt{3}$. In this regime, the wall velocity does not have a strong impact on the GW peak amplitude. For wall velocities smaller than the speed of sound, the efficiency factor decreases rapidly and the generation of GW from sound waves is suppressed~\cite{Cutting:2019zws}.

\subsubsection{Efficiency factors}
\label{sec:eff-factors}
The efficiency factors determine which fraction of the energy budget is converted into GWs. In this work, we focus on the GWs from sound waves, which is the dominating contribution for the phase transitions considered here. The efficiency factor for the sound waves $\kappa_\text{sw}$ consist of the factor $\kappa_v$ \cite{Espinosa:2010hh} as well as an additional suppression due to the length of the sound-wave period $\tau_\text{sw}$ \cite{Ellis:2019oqb, Ellis:2020awk, Guo:2020grp}
\begin{align}
	\kappa_\text{sw}&= \sqrt{\tau_\text{sw}} \, \kappa_v\,.\label{eq:efficiency}
\end{align}
In our notation, $\tau_\text{sw}$ is dimensionless and measured in units of the Hubble time. It is given by \cite{Guo:2020grp}
\begin{align}
	\tau_\text{sw}=1-1/\sqrt{1+2\frac{(8\pi)^{\frac 13} v_w}{\tilde \beta \,\bar U_f}}\,. \label{eq:suppression}
\end{align}
where $\bar U_f$ is the root-mean-square fluid velocity \cite{Hindmarsh:2015qta, Ellis:2019oqb}
\begin{align}
	\bar U_f^2 = \frac{3}{v_w(1+\alpha)}\int^{v_w}_{c_s}\!\mathrm d\xi \, \xi^2 \frac{v(\xi)^2}{1-v(\xi)^2}\simeq \frac{3}{4}\frac{\alpha}{1+\alpha}\kappa_v \,.
\end{align}
We follow \cite{Espinosa:2010hh} for $\kappa_v$ where it was numerically fitted to simulation results. The factor $\kappa_v$ depends $\alpha$ and $v_w$, and, for example, at the Chapman-Jouguet detonation velocity it reads
\begin{align}
	\kappa_v(v_w=v_J)=\frac{\sqrt{\alpha}}{0.135 +\sqrt{0.98+\alpha}}\,.
	\label{eq:eff-vJ}
\end{align}
For the confinement phase transition with $\alpha \approx 1/3$, this leads to $\kappa_v\approx 0.45$, and for the chiral phase transition with $\alpha \sim \mathcal{O}(10^{-2})$, we have $\kappa_v \sim 0.1$.

\begin{table*}[t!]
	\resizebox{\textwidth}{15mm}{
		\begin{tabular}{|c||c|c|c||c|c||c|c|c|}
		\hline
		Model & \multicolumn{3}{c||}{PNJL parameters} & \multicolumn{2}{c||}{\;\;GW parameters\;\;} & \multicolumn{3}{c|}{Observables} \\
		\hline
		&\;\;$G_S\,[\text{GeV}^{-2}]$\;\; & \;\;$G_D\,[\text{GeV}^{-5}]$\;\;&\;\; $\Lambda/T_0$\;\; &$\alpha$ & $\tilde \beta/10^4$ & \;\;$M\,[\text{GeV}]$\;\; & \;\;$f_\pi\,[\text{GeV}]$\;\; & \;$m_\sigma\,[\text{GeV}]$\;\\
		\hline
		\;\;3F3 (benchmark)\;\; &4.6	&-743	& 3.54	&\;\;\,0.029\;\;\,	&1.9	&203	&63		&416\\
		\hline
		3F3 (best case)		 &4.6	&-1486	 & 3.54	&0.051			&0.68	&260	&60		&526\\
		\hline
		3G1 (benchmark)	  &7.5	  &0	 	  & 7.37	&0.34			&8.1	&1261	&183		&2523\\
		\hline
		3G1 (best case) 	&7.5	&0			& 14.7 	& 0.34			 & 7.6 	&10917	&198	&21834\\
		\hline
		3S1 (benchmark)	  &44.4	 &0			 & 5.70    & 0.34 		&5.9	&1105	&118	&2210\\
		\hline
		3S1 (best case) 	&44.4	&0			& 3.35	& 0.36			&1.7	&70		&45		&140\\
		\hline
	\end{tabular}}
	\caption{\label{tab:GW-param-obs}
		Table of parameters of the PNJL model with resulting GW parameters and observables for the benchmark and best-case scenario for $T_c=100$\,GeV.
	}
\end{table*}

\subsection{Gravitational-wave spectrum}
\label{sec:spectrum}
We follow the treatment in \cite{Caprini:2015zlo, Caprini:2019egz} for extracting the GW spectrum from the parameters $\alpha$, $\tilde \beta$, and $v_w$. We focus on the contribution from sound waves in the plasma after bubble collision \cite{Hindmarsh:2013xza, Giblin:2013kea, Giblin:2014qia, Hindmarsh:2015qta, Hindmarsh:2017gnf}. The contributions from bubble collision \cite{Kosowsky:1991ua, Kosowsky:1992rz, Kosowsky:1992vn, Kamionkowski:1993fg, Caprini:2007xq, Huber:2008hg, Caprini:2009fx, Espinosa:2010hh, Weir:2016tov, Jinno:2016vai} and magnetohydrodynamic turbulence in the plasma \cite{Kosowsky:2001xp, Dolgov:2002ra, Caprini:2006jb, Gogoberidze:2007an, Kahniashvili:2008pe, Kahniashvili:2009mf, Caprini:2009yp, Kisslinger:2015hua} are subleading. The latter will be explicitly shown. The GW spectrum from sound waves is given by
\begin{align}
	h^2\Omega_\text{GW}(f)&= h^2\Omega^\text{peak}_\text{GW} \left(\frac{f}{f_\text{peak}}\right)^{\!3} \left[ \frac{4}{7}+\frac{3}{7}\left( \frac{f}{f_\text{peak}} \right)^{\!2}\right]^{-\frac{7}{2}}\!,
	\label{eq:GWsignal}
\end{align}
with the peak frequency
\begin{align}
	f_\text{peak}&\simeq 1.9\cdot 10^{-5}\,\text{Hz}\left(\frac{g_*}{100} \right)^{\!\frac{1}{6}}\left( \frac{T}{100\, \text{GeV}}\right) \left(\frac{\tilde \beta}{v_w} \right),
	\label{eq:peak-f}
\end{align}
and the peak amplitude
\begin{align}\label{eq:peak-amp}
	h^2\Omega^\text{peak}_\text{GW} &\simeq 2.65\cdot 10^{-6}\left(\frac{v_w}{\tilde \beta}\right)\left( \frac{\kappa\, \alpha}{1+\alpha} \right)^{\!2}\left(\frac{100}{g_*}\right)^{\!\frac{1}{3}}\Omega_\text{dark}^2\,.
\end{align}
Here, $h= H/(100 \text{km}/\text{s}/\text{Mpc})$ is the dimensionless Hubble parameter and $g_*$ is the effective number of relativistic d.o.f., including the the SM d.o.f.\ $g_{*,\text{SM}}=106.75$ and the dark sector ones, which is $g_{*,\text{dark}} = 47.5$ in the fundamental case and $g_{*,\text{dark}} = 17$ in the adjoint and two-index symmetric case. The latter stems from 16 gluonic and 1 dark pion d.o.f..

The factor $\Omega_\text{dark}^2$ in \cref{eq:peak-amp} accounts for the dilution of the GWs by the visible SM matter which does not participate in the phase transition. The factor reads
\begin{align}
	\label{eq:dilution}
	\Omega_\text{dark} =\frac{\rho_{\text{rad},\text{dark}}}{\rho_\text{rad,tot}}=\frac{g_{*,\text{dark}} }{g_{*,\text{dark}}+g_{*,\text{SM}}}\,.
\end{align}
The decisive quantity that determines the detectability of a GW signal is the signal-to-noise-ratio (SNR) \cite{Allen:1997ad, Maggiore:1999vm} given by
\begin{align}
	\text{SNR} = \sqrt{\frac{T}{\text{s}} \int_{f_\text{min}}^{f_\text{max}}\mathrm df \left(\frac{ h^2 \Omega_\text{GW}}{h^2 \Omega_\text{det}} \right)^{\!2}}.
	\label{eq:SNR}
\end{align}
Here, $h^2 \Omega_\text{GW}$ is the GW spectrum given by \cref{eq:GWsignal}, $h^2 \Omega_\text{det}$ the sensitivity curve of the detector, and $T$ the observation time, for which we assume $T=3$\,years. We compute the SNR of the GW signals for the future GW observatories LISA \cite{Audley:2017drz, Baker:2019nia, LISAdocument}, BBO \cite{Crowder:2005nr, Corbin:2005ny, Harry:2006fi, Thrane:2013oya, Yagi:2011wg}, and DECIGO \cite{Seto:2001qf,Yagi:2011wg, Kawamura:2006up, Isoyama:2018rjb}. The sensitivity curves of these detectors are nicely summarised and provided in \cite{Schmitz:2020syl}.

\section{Results}
\label{sec:results}
With the setup of the PNJL model in \cref{sec:eff-theories} and the tools for phase transition in \cref{sec:phase-transition}, we are now ready to display our results.

\subsection{Choice of PNJL parameters}
The fundamental QCD-type model of our dark gauge-fermion sector has only one parameter, which can be for example the critical temperature $T_c$ or the value of the strong coupling at any arbitrary scale. However, we are working in the effective PNJL model which has the advantage of being well suited to describe the strong dynamics at the phase transition, but also has the drawback of many model parameters, including the couplings $G_S$ and $G_D$, the cutoff $\Lambda$, and the $a_i,\, b_i$ coefficients as well as the temperature $T_0$ in the Polyakov-loop potential. The $a_i,\, b_i$ coefficients have been fitted against pure-glue lattice data, see \cref{tab:best-fit}, and we make the approximation that the coefficients remain the same in the presence of quarks.

In the 3F3 model, we have further guidance from the lattice and we use values from \cite{Fukushima:2017csk} as a benchmark, and rescale them such that, for example, $T_c = 100$\,GeV, see \cref{tab:GW-param-obs}. We also scan the PNJL parameters in the vicinity of this benchmark point and search for a best-case scenario that gives us the strongest GW signal and thereby explore the lattice uncertainty on the PNJL model parameters. Note that we keep the ratio $\Lambda/T_0 = 3.8$ fixed such that it agrees with \cite{Fukushima:2008wg}.

\begin{figure*}[t]
	\includegraphics[width=0.99\linewidth]{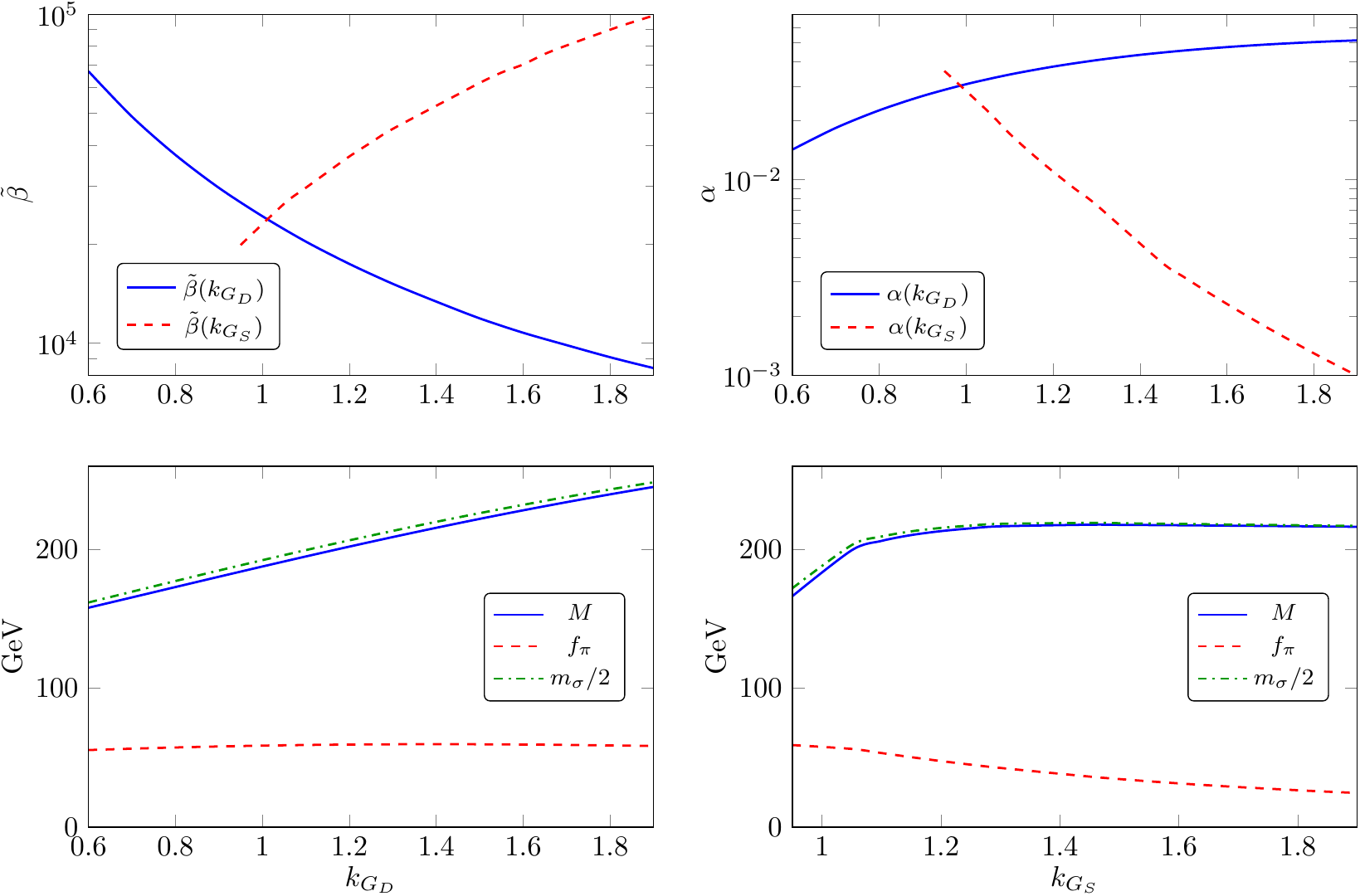}
	\caption{Fundamental representation: The GW parameters $\tilde \beta$ (top left) and $\alpha$ (top right) as well as the observables $M$, $f_{\pi}$, and $m_\sigma$ (bottom) as a function of the couplings $G_S = k_{G_S}\cdot 4.6\,\text{GeV}^{-2}$ and $G_D = k_{G_D}\cdot(-743\,\text{GeV}^{-5})$. We use $T_c=100$\,GeV, the ratio $\Lambda/T_0 = 3.54$, as well as the polynomial parameterization of the Polyakov-loop potential. Below $k_{G_S,\text{crit}} = 0.882$, no chiral symmetry breaking occurs.}
	\label{fig:fund}
\end{figure*}

In the 3G1 and 3S1 models, there are no lattice data available\footnote{There are quite a few lattice works in the direction of fermions with higher dimension representations. However, most of these focus on studying the phase structure between chiral symmetry breaking and (near)conformal phase (see e.g.~\cite{Karsch:1998qj, Athenodorou:2014eua}) rather than producing the mass spectrums which are important in determining the PNJL parameters.} to fit to as in the 3F3 case and therefore the PNJL parameters are basically unconstrained. Nonetheless, we use the benchmark values\footnote{These values are obtained based on the parameter values in the fundamental representations and through the group theoretical transformation. This is built upon the assumption that the NJL model effective coupling in any representations can be linked to QCD fundamental gauge coupling through a Fierz transformation where only group theoretical factors are involved \cite{Zhang:2010kn}.} from \cite{Kahara:2012yr}, again rescaled to, e.g., $T_c = 100$\,GeV, see \cref{tab:GW-param-obs}. We again scan the parameters for the best-case scenario, however, in these models we also vary the ratio $\Lambda/T_0$ due to the lack of constraints from the lattice data. Remarkably, although the parameters have few constraints, we find clear predictions for these models.

\subsection{Parameter scan}
\label{sec:param-scan}
\textbf{Fundamental Representation:}
We vary the couplings $G_S$ and $G_D$, while we keep the ratio $\Lambda/T_0$ fixed, since the latter is well constrained by lattice data. We parameterize the couplings with $G_S = k_{G_S}\cdot 4.6\,\text{GeV}^{-2}$ and $G_D = k_{G_D}\cdot(-743\,\text{GeV}^{-5})$, where $k_{G_S} = k_{G_D} = 1$ is the benchmark point from \cite{Fukushima:2017csk}, which we choose as a starting point of the scan. Below a certain value of $k_{G_S}$, chiral symmetry breaking does not occur. This value depends on $k_{G_D}$ and, for example, for $k_{G_D}=1$, we have $k_{G_S,\text{crit}} = 0.882$.

The resulting GW parameters and PNJL observables are displayed in \cref{fig:fund}. For the purpose of the plot, we keep either $k_{G_S}$ or $k_{G_D}$ equal to one, while varying the other. We observe that the strength of the phase transition is increasing with a larger $G_D$ and decreasing with larger $G_S$, i.e., $\tilde \beta$ is decreasing and $\alpha$ is increasing with a larger $G_D$. The phase transition cannot become arbitrarily strong, instead the GW parameters approach asymptotic values which we estimate via extrapolation to be $\tilde \beta_\text{asymp} \approx 3.5\cdot 10^3$ and $\alpha_\text{asymp} \approx 0.06$.
Consequently, we use a large value of $G_D$ as best-case scenario in \cref{tab:GW-param-obs}.

In terms of PNJL observables, we observe that the constituent mass $M$ and the sigma-meson mass $m_\sigma$ grow roughly linear with $G_S$ and $G_D$, while the pion decay constant remains roughly constant. Also we observe the approximate relation $2 M \approx m_\sigma$.

\begin{figure*}[t]
	\includegraphics[width=0.99\linewidth]{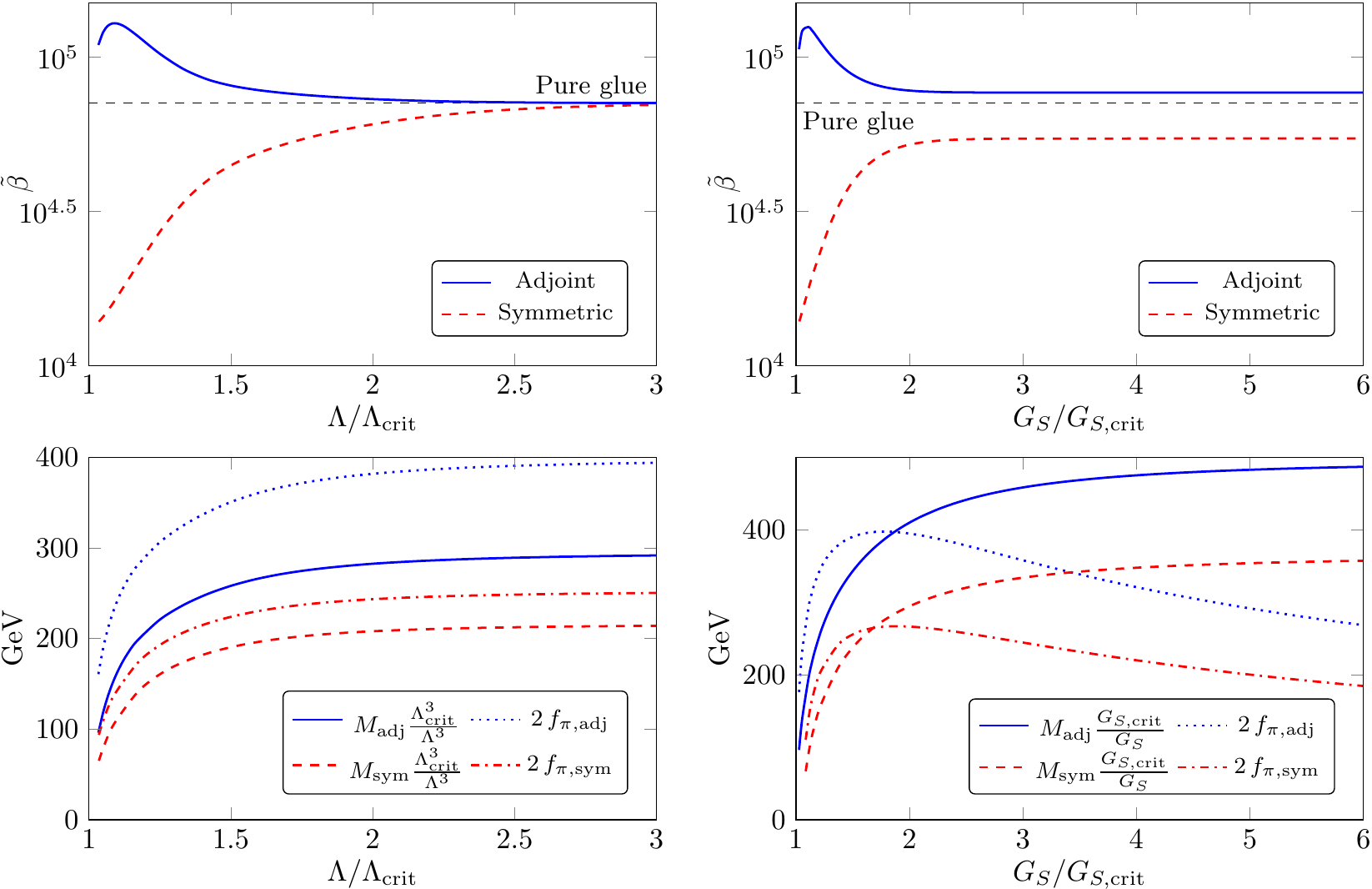}
	\caption{The GW parameter $\tilde \beta$ (top) and the PNJL observables $M$ and $f_{\pi}$ (bottom) as a function of the cutoff $\Lambda$ (left) and the coupling $G_S$ (right) with fermions in the adjoint and two-index symmetric representations. We use $T_c=100$\,GeV and the polynomial parameterization of the Polyakov-loop potential.}
	\label{fig:confinement-all}
\end{figure*}

\textbf{Adjoint and two-index symmetric Representation:} We vary the coupling $G_S$ and the cutoff $\Lambda$ while the temperature $T_0$ is determined by the choice of $T_c$. We start from the benchmark values in \cite{Kahara:2012yr}, see \cref{tab:GW-param-obs}. The resulting GW parameters and PNJL observables are displayed in \cref{fig:confinement-all}. For the purpose of the plot, we keep one coupling fixed to the benchmark value while varying the other.

Both, $G_S$ and $\Lambda$, have a lower bound below which chiral symmetry breaking does not occur. We denote these lower bounds by $G_{S,\text{crit}}$ and $\Lambda_\text{crit}$. If the respectively other coupling is at the benchmark point, they take the values $G_{S,\text{crit,adj}} = 2.68\,\text{GeV}^{-2}$ and $\Lambda_\text{crit,adj}/T_0= 4.41$ in the adjoint representation case as well as $G_{S,\text{crit,sym}} = 14.3\,\text{GeV}^{-2}$ and $\Lambda_\text{crit,sym}/T_0= 3.24$ in the two-index symmetric representation case.

For $\Lambda/\Lambda_\text{crit}\gg1$, the fermions are decoupling from the theory and the pure-glue result is approached, see \cref{fig:confinement-all}. This is most easily understood in terms of the constituent mass, which is growing with $\Lambda^3$ and triggers this decoupling. The large constituent mass makes the medium potential trivial, see \cref{eq:medium-pot}, and we are left with the pure-glue dynamics.

For $\Lambda/\Lambda_\text{crit}\to 1$, we observe that the fermion representation makes a big difference for the GW parameter $\tilde \beta$. Fermions in the adjoint representation lead to larger values of $\tilde \beta$ (weaker phase transition), while fermions in the two-index symmetric representation lead to smaller values of $\tilde \beta$ (stronger phase transition).

Also when varying the parameter $G_S$, we observe that the two-index symmetric representation case always has smaller values of $\tilde\beta$ than the pure-glue case while the adjoint representation case always has larger values of $\tilde\beta$. In terms of the PNJL observables, we see that the constituent mass grows linear with $G_S$ while the pion decay constant is decreasing with $G_S$. For $G_S/G_{S,\text{crit}}\gg 1$, the GW parameter $\tilde \beta$ becomes constant but does not approach the pure-glue result, although the constituent mass is growing linear with $G_S$. The difference to the large-$\Lambda$ limit is that we implement a momentum-dependent four-fermion coupling, see \cref{eq:mom-dep-cutoff}, and therefore the mediums potential still gives contributions for momenta larger than the cutoff.

Consequently, we choose for the best-case scenario a small value of the cutoff in the two-index symmetric representation case and a large value of the cutoff in the adjoint representation case, see \cref{tab:GW-param-obs}.

\subsection{Gravitational-wave spectrum}
\label{eq:res-GW-spec}
With the GW parameters displayed in \cref{tab:GW-param-obs}, we compute the GW spectrum as described in \cref{sec:spectrum}. We use both, the polynomial and the logarithmic parameterization of the Polyakov-loop potential, see \cref{eq:PLM_potential,eq:PLM_SU3_Log}. The central value of both parameterizations is our main result, $\beta  = (\beta_\text{poly}+\beta_\text{log})/2$, and we estimate the error of the parameters via the difference between the parameterizations, $\delta \beta = |\beta_\text{poly}-\beta_\text{log}|/2$. Furthermore we vary the wall velocity between the speed of sound and light speed, $c_s \leq v_w \leq 1$. With the latter treatment, we define the error bands for GW signal. This is compared to the power-law integrated sensitivity curves of LISA \cite{Audley:2017drz, Baker:2019nia, LISAdocument}, BBO \cite{Crowder:2005nr, Corbin:2005ny, Harry:2006fi,Thrane:2013oya, Yagi:2011wg}, and DECIGO \cite{Seto:2001qf, Yagi:2011wg, Kawamura:2006up, Isoyama:2018rjb}. The power-law integrated sensitivity curves provide a qualitative visualisation for the detectability of a GW signal. Since the GW signals considered here are not simple power-law signals within the frequency range of the detector, we refer to the SNR for the quantitative discussion of detectability of the GWs, which is discussed in \cref{sec:SNR}.

\begin{figure}[t]
	\includegraphics[width=\linewidth]{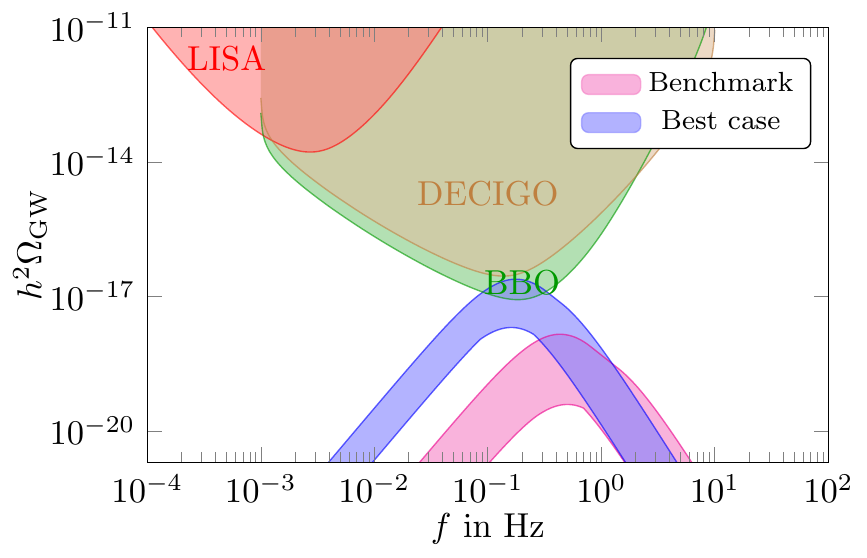}
	\caption{Gravitational-wave spectrum with three Dirac fermions in the fundamental representation for $T_c=100$\,GeV.
	}
	\label{fig:GW-Fund}
\end{figure}

\textbf{Fundamental Representation:}
We present the GW spectrum from the chiral phase transition with fermions in the fundamental representation in \cref{fig:GW-Fund}. The best-case scenario has a peak frequency of $h^2\Omega^\text{peak}_\text{GW}\sim \mathcal{O}(10^{-17})$ and might be detectable with the BBO. The benchmark scenario has a peak frequency of $h^2\Omega^\text{peak}_\text{GW}\sim \mathcal{O}(10^{-18})$ and is out of reach of any planned future detector.

\begin{figure}[b]
	\includegraphics[width=\linewidth]{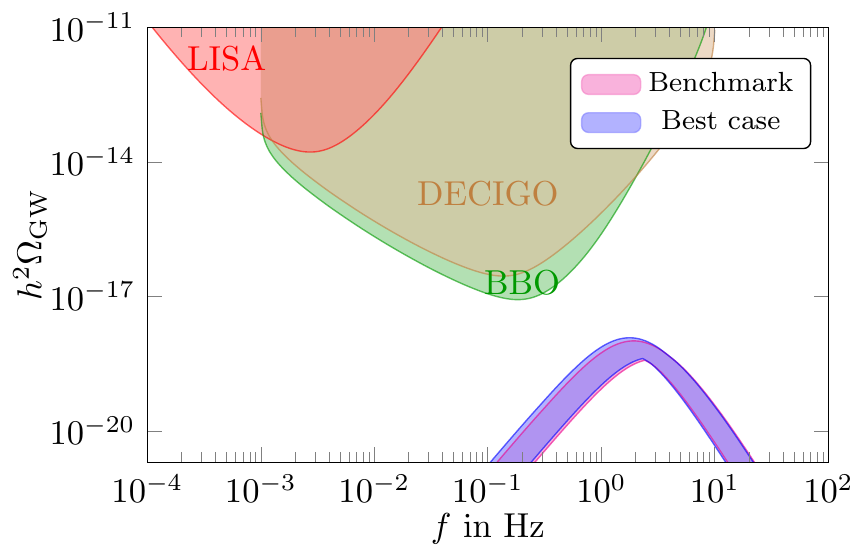}
	\caption{Gravitational-wave spectrum with one Dirac fermion in the adjoint representation for $T_c=100$\,GeV.
	}
	\label{fig:GW-Adjoint}
\end{figure}

\textbf{Adjoint Representation:}
We present the GW spectrum from the confinement phase transition with fermions in the adjoint representation in \cref{fig:GW-Adjoint}. The GW signal is strongly suppressed and out of reach of any planned future detector. It has a peak frequency of $h^2\Omega^\text{peak}_\text{GW}\sim \mathcal{O}(10^{-18})$. Remarkably, the spectra from the benchmark and the best-case scenario are almost identical, which shows that in this case, the PNJL model is highly predictive.

\begin{figure}[t]
	\includegraphics[width=\linewidth]{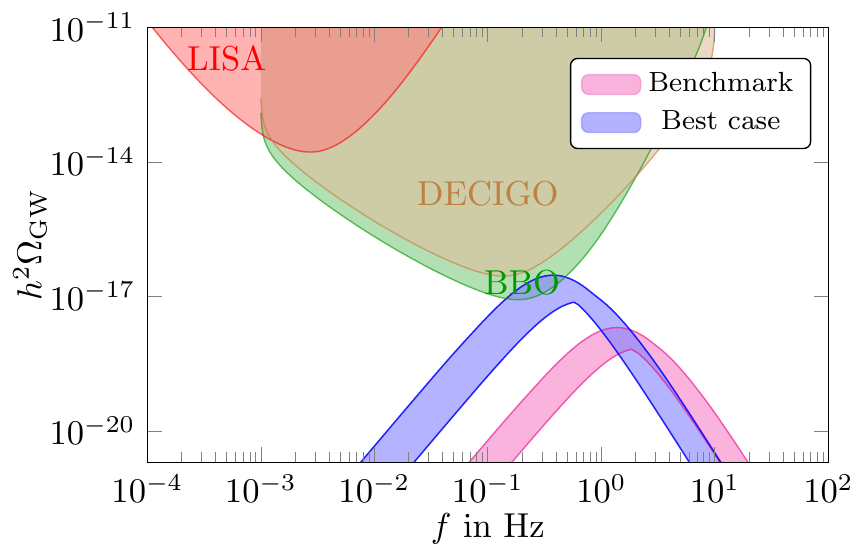}
	\caption{Gravitational-wave spectrum with one Dirac fermion in the two-index symmetric representation for $T_c=100$\,GeV.
	}
	\label{fig:GW-Sym}
\end{figure}

\textbf{Two-index symmetric Representation:}
We present the GW spectrum from the confinement phase transition with fermions in the two-index symmetric representation in \cref{fig:GW-Sym}. Similar to the GW spectrum of the fundamental case, the best-case scenario has a peak frequency of $h^2\Omega^\text{peak}_\text{GW}\sim \mathcal{O}(10^{-17})$ and might be detectable with the BBO, while the benchmark scenario has a peak frequency of $h^2\Omega^\text{peak}_\text{GW}\sim \mathcal{O}(10^{-18})$ and is out of reach of any planned future detector.

\begin{figure}[b]
	\includegraphics[width=\linewidth]{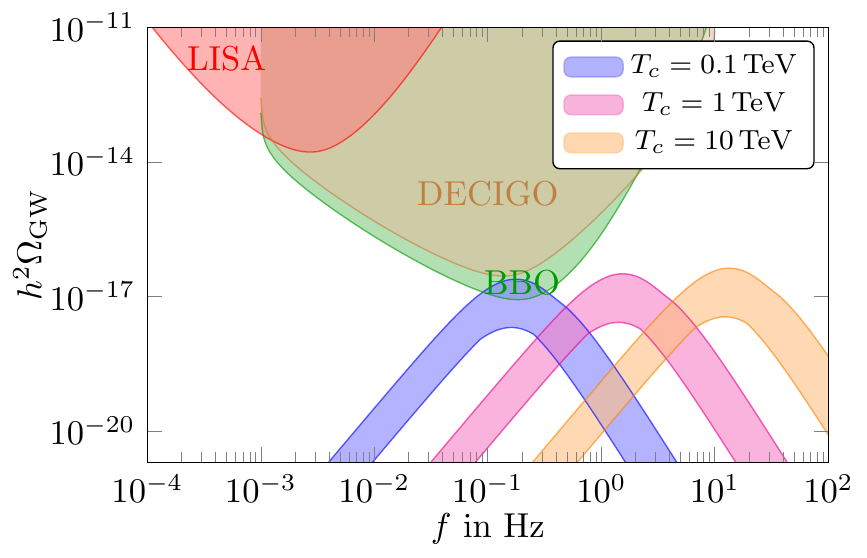}
	\caption{Gravitational-wave spectrum with three Dirac fermions in the fundamental representation for different critical temperatures.
	}
	\label{fig:GW-Fund-Tc}
\end{figure}

\textbf{Dependence on $T_c$:}
In \cref{fig:GW-Fund-Tc}, we show the dependence of the GW spectrum on the critical temperature at the example of the best-case scenario of the fundamental representation. As expected, the critical temperature simply shifts the peak frequency of the GW spectrum. For the fundamental representation, we have the biggest overlap with BBO for $T_c\sim 100$\,GeV. For the two-index symmetric representation, this happens at  $T_c\sim 50$\,GeV, and for the adjoint representation even at $T_c\sim 10$\,GeV, see also \cref{sec:SNR}.

\begin{figure}[t]
	\includegraphics[width=\linewidth]{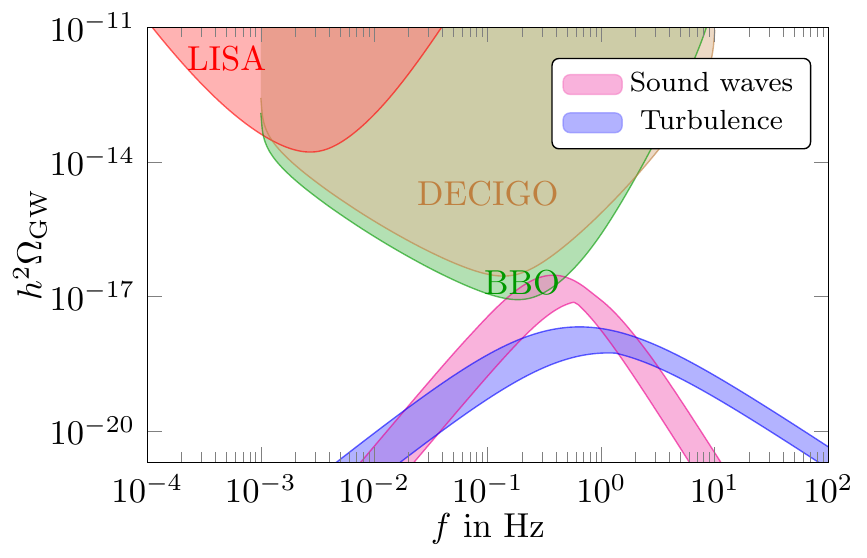}
	\caption{Gravitational-wave spectrum from sound waves in comparison with the contribution from turbulence at the example of the two-index symmetric representation for $T_c=100$\,GeV.
	}
	\label{fig:GW-Turbulence}
\end{figure}

\textbf{Sound waves vs turbulence:}
In \cref{fig:GW-Turbulence}, we display the comparison between GWs from sound waves and from turbulence at the example of the best-case scenario of the two-index symmetric representation. The sound-wave contribution is clearly dominating over the turbulence with the exception of frequencies far above and below the peak frequency. This is related to the slower fall-off behaviour of the turbulence contribution compared to the sound-wave contribution. We have checked that the sound waves are dominating the GW spectra for all phase transitions considered here.

\subsection{Signal-to-noise ratio}
\label{sec:SNR}
\begin{figure}[t]
	\includegraphics[width=\linewidth]{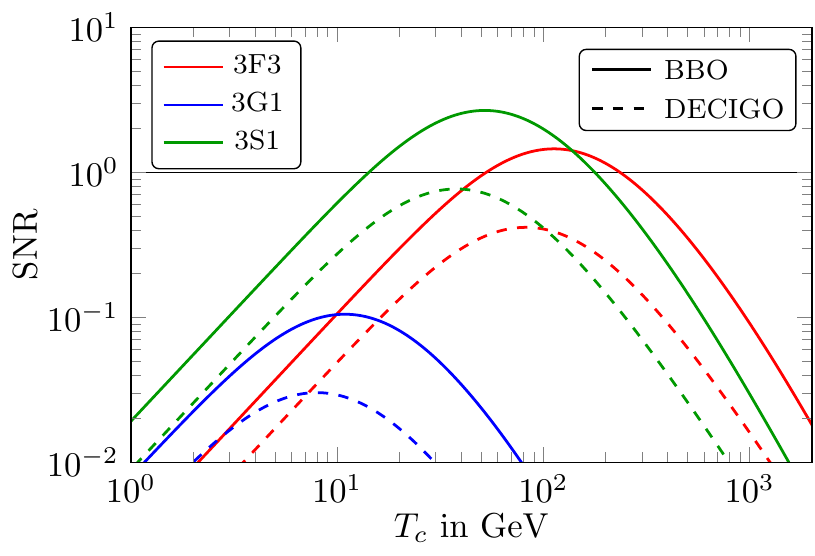}
	\caption{Signal-to-noise ratio as a function of the critical temperature for the best-case scenarios of each model at BBO and DECIGO. We assumed an observation time of three years.
	}
	\label{fig:SNR}
\end{figure}

In \cref{fig:SNR} we present the signal-to-noise ratio for the detectors BBO and DECIGO as a function of the critical temperature for the best-case scenario of the three models, see \cref{tab:GW-param-obs}. The SNR of the LISA detector is too small to be displayed in the plot. For all SNRs, we assumed an observation time of three years, see \cref{eq:SNR}.

If we assume that for a successful detection we need $\text{SNR}>1$\footnote{In general, it is difficult to determine the exact threshold from which a signal is detectable. It also depends on how well the astrophysical foreground such as gravitational radiation from inspiralling compact binaries can be subtracted from the signal, see, e.g., \cite{Cutler:2005qq, Pan:2019uyn, Lewicki:2021kmu}. Here we make the probably optimistic assumption that a signal with $\text{SNR}>1$ is detectable.}, then BBO will test the chiral phase transition in the 3F3 model for $50\,\text{GeV}\lesssim T_c \lesssim 200\,\text{GeV}$ as well as the confinement phase transition in the 3S1 model for $10\,\text{GeV}\lesssim T_c \lesssim 200\,\text{GeV}$. The SNR at DECIGO is slightly below one for these models for all $T_c$. A detection of the chiral phase transition in the 3G1 model is out of reach for both detectors.

\section{Discussions}
\label{sec:discussion}

\subsection{Thin-wall Approximation}
All phase transitions in the gauge-fermion systems studied here, as well as the phase transitions in $SU(N)$ gauge theories without fermions studied in \cite{Huang:2020crf}, have one thing in common: the inverse duration is remarkably large $\tilde \beta \sim \mathcal{O}(10^4)$. Here we want to provide an instructive argument via the thin-wall approximation to explain this feature in an intuitive picture. The advantage of the thin-wall approximation is that we can analytically calculate the decay rate of the false vacuum in terms of the latent heat and the surface tension. Eventually, we will relate the large inverse duration time to a competition between the latent heat and the surface tension.

The thin-wall approximation for the Euclidean action was derived in \cite{Linde:1981zj, Fuller:1987ue} and we briefly review it here. The three-dimensional Euclidean action is written as
\begin{align}
	S_3=\frac{4\pi}{3}r_c^3\left(p_\text{f}-p_\text{t}\right)+4\pi\sigma r_c^2\,,
	\label{eq:S_3}
\end{align}
where $p_\text{f}$ and $p_\text{t}$ denote respectively the pressure in the false vacuum and true vacuum, $\sigma$ is the surface tension of the nucleation bubble, and $r_c$ is the critical radius of the nucleation bubble defined by
\begin{align}
	p_\text{t}-p_\text{f}=\frac{2\sigma}{r_c}\,.
	\label{eq:surface_tension}
\end{align}
On the other hand, the difference in the pressure between the false vacuum and true vacuum is also linked to the latent heat $L$ via
\begin{align}
	p_\text{t}-p_\text{f}&=L\eta\,,
	&
	\text{with}\quad \eta&=\frac{T_c-T}{T_c}\,.
	\label{eq:Latent_Heat}
\end{align}
The thin-wall approximation works with the assumption that $\eta$ is small, $\eta \ll 1$. With \cref{eq:surface_tension,eq:Latent_Heat}, the three-dimensional Euclidean action \cref{eq:S_3} can be written as a function of the latent heat $L$ and the surface tension $\sigma$
\begin{align}
	S_3=\frac{16\pi}{3}\frac{\sigma(T_c)^3}{L(T_c)^2} \frac{T_c^2}{(T_c-T)^2}\, ,
	\label{eq:S3_thin-wall}
\end{align}
where we have made explicit that the surface tension and latent heat are evaluated at $T_c$. The ratio $S_3(T_p)/T_p$ is typically a number $\mathcal{O}(150)$ for phase transitions around the electroweak scale. From this we infer that
\begin{align}
	\label{eq:Tp_thin-wall}
	T_c- T_p \approx \sqrt{\frac{16 \pi \sigma^3 T_c}{3L^2\cdot\mathcal{O}(150) }}\,,
\end{align}
and the inverse duration $\tilde{\beta}$ follows as
\begin{align}
	\tilde{\beta}&=T\frac{\mathrm d}{\mathrm dT}\frac{S_3(T)}{T}\bigg|_{T=T_p}\approx \mathcal{O}(10^3) \frac{T_c^{1/2} L }{\sigma^{3/2}} \,,
	\label{eq:thin_wall_beta}
\end{align}
see also \cite{Eichhorn:2020upj}. We see that $\tilde{\beta}$ stems from the competition between latent heat $L$ and surface tension $\sigma$. However, already the prefactor is $\mathcal{O}(10^3)$ and therefore we would need a surface tension that is much larger than the latent heat to achieve a strong first-order phase transition. We caution that this analysis relies on $\eta \ll 1$, which is true for the phase transitions investigated here but must not hold for all strongly coupled gauge-fermion systems.

\begin{table*}[t!]
	\resizebox{150mm}{30mm}{
	\begin{tabular}{|c|c|c|c|c|c|}
		\hline
		\;Model name\; & Gauge group & \;Fermion irrep\; & \;Reality\; & \;Number of Weyl flavours\; & \;Centre Symmetry\; \\
		\hline
		M1 & $SU(N),\;N\geq 3$ & $\textbf{F}$ & C & 6 & $\varnothing$ \\
		\hline
		M2 & $Sp(2N),\;N\geq 1$ & $\textbf{F}$ & PR & 6 & $\varnothing$ \\
		\hline
		M3 & $SU(4)$ & $\textbf{A}_2$ & R & 3 & $Z_2$ \\
		\hline
		M4 & $Sp(4)$ & $\textbf{A}_2$ & R & 3 & $Z_2$ \\
		\hline
		M5 & \;$Spin(N),\; N\geq 7$\; & $\textbf{F}$ & R & 3 & $Z_2$ \\
		\hline
		M6 & $Spin(8)$ & $\textbf{Spin}$ & R & 3 & $Z_2$ \\
		\hline
		M7 & $G_2$ & $\textbf{F}$ & R & 3 & $\varnothing$ \\
		\hline
		M8 & $SU(5)$ & $\textbf{A}_2$ & C & 2 & $\varnothing$ \\
		\hline
		M9 & $SU(3)$ & $\textbf{Adj}$ & R & 1 & $Z_3$ \\
		\hline
		M10 & $Sp(4)$ & $\textbf{Adj}$ & R & 1 & $Z_2$ \\
		\hline
		M11 & $Sp(8)$ & $\textbf{A}_2$ & R & 1 & $Z_2$ \\
		\hline
		M12 & $F_4$ & $\textbf{F}$ & R & 1 & $\varnothing$ \\
		\hline
	\end{tabular}}
	\caption{\label{tab:cubicmodels}
		List of gauge-fermion models that lead to cubic terms in the condensate energy of the corresponding NJL model.
	}
\end{table*}

In the case of pure-gluon dynamics, lattice results have provided fitting functions for the surface tension and the latent heat as functions of the number of colours and $T_c$ \cite{Panero:2009tv, Lucini:2005vg}. With those lattice results, we find the small surface tension can not compensate the large latent heat and thus leads to $\tilde{\beta}\sim \mathcal{O}(10^4)$ \cite{Huang:2020crf}. We expect that it is quite common that in strongly coupled gauge-fermion systems, the latent heat is big while the surface tension is not sufficiently large and thus many small bubbles are formed (nucleate very quickly everywhere in the space-time) rather than large and long-lasting big bubbles which are crucial to generate a strong GW signal.

\subsection{Models with Cubic Terms in the Condensate Energy}
In the three gauge-fermion models we explored, a first-order chiral phase transition is only found for the 3F3 case which features a cubic ($\signob^3$) term in its condensate energy, see \cref{eq:vcond3F3}. For the 3G1 and 3S1 case such a cubic term is absent and we do not discover a first-order chiral phase transition in the parameter space. The relevance of the cubic term for the triggering of a first-order chiral phase transition is well-known~\cite{Fukushima:2013rx}. In the 3F3 case, such a cubic term originates from the 't Hooft determinantal interaction. It is therefore tempting to ask the following question: Suppose we consider a gauge-fermion theory featuring a simple gauge group and one type of fermion under a single irreducible representation of the gauge group (for complex representations, its conjugate does not count), can we list all such gauge-fermion models that deliver a six-fermion 't Hooft determinantal term so that their NJL condensate energy contain cubic terms? The answer is positive. It turns out there is a finite number of such models, characterized by the condition coming from an analysis of the discrete chiral symmetry that must be preserved by the 't Hooft determinantal term
\begin{align}
	2N_f^W T(r)=6 \,,
\end{align}
with $N_f^W$ being the number of Weyl flavours, $T(r)$ is the trace normalization factor for the fermion representation $r$. The resulting solutions are listed in \cref{tab:cubicmodels}, in which we also list the remnant centre symmetry for each case. The 3F3 model belongs to the M1 category. The models listed in the table are the natural targets for the next-step investigation. They require extensions of the effective theory framework used in this work so that larger gauge groups can be dealt with, as commented in \cref{subsec:2E}.

The possibility of having both a first-order chiral phase transition and a first-order confinement phase transition is intriguing. If they are separated in scale, a double-peak feature in the GW signature is in principle possible. \cref{tab:cubicmodels} gives some hints on where to find candidate models with this feature. Here actually the requirement of centre symmetry is not that strict. If a nontrivial centre symmetry remains, one may have an idea about the order of the confinement phase transition from universality argument~\cite{Fukushima:2013rx}. However, as commented in \cref{subsec:2G}, only the absence of a FP requires a first-order phase transition, while the existence of a FP does not lead to definite predictions. Therefore, although an $SU(2)$ confinement phase transition (with $Z_2$ centre) is known to be of second-order, it is not straightforward to exclude the possibility of first-order phase transitions for other cases with $Z_2$ centre listed in \cref{tab:cubicmodels}. Moreover, even if the remnant centre symmetry is trivial, there still can be a first-order confinement phase transition. There can be two possibilities in such a case. The first is that the centre symmetry is only broken weakly by dynamical quarks. This is the case when the chiral phase transition occurs at a scale a few times larger than the confinement transition scale (which is typical in higher representations, see \cite{Evans:2020ztq}). When the temperature drops below the chiral phase-transition temperature, the quarks obtain a large dynamical mass and are decoupled from the low-energy gauge dynamics. A first-order confinement phase transition can then occur as if the quarks are absent. In fact, this phenomenon already takes place in the 3S1 case we studied and has been found earlier in \cite{Kahara:2012yr}. The second is that the confinement phase transition might only be directly related to the change of some order parameter, but not related to the centre symmetry. For example, the $G_2$ gauge group is known to have a trivial centre even in the absence of dynamical quarks, but lattice studies suggest it exhibits a first-order confinement phase transition at finite temperature~\cite{Pepe:2006er, Cossu:2007dk}. Modelling the confinement phase transitions in such cases is an interesting issue which we leave for future work. Finally, we would like to comment that the model M9 corresponds to the supersymmetric $SU(3)$ gauge theory. A recent lattice study~\cite{Bergner:2019kub} suggests that it exhibits a single first-order phase transition where chiral symmetry is restored and centre symmetry gets broken at high temperature.

\section{Conclusions}
\label{sec:conclusions}
We studied the GW signal from strongly coupled gauge-fermion systems featuring both, dark chiral and confinement phase transitions. We employed the PNJL model to systematically study $SU(3)$ Yang-Mills theory with fermions in fundamental, adjoint, and two-index symmetric representations. We studied in detail the interplay between chiral and confinement phase transitions.

We discovered that the representation of the fermions matters: the two-index symmetric representation case leads to the strongest first-order phase transition and has the highest chance of being detected by the Big Bang Observer experiment with a potential signal-to-noise ratio of $\text{SNR}\sim 10$. Conversely, fermions in the adjoint representation lead to the weakest GW signal.  However, for all models considered here, the inverse duration time is large, $\tilde \beta \sim \mathcal{O}(10^4)$, and the GW signal is generically suppressed. We analyse this observation through the thin-wall approximation and show that the large rate stems from the competition between the small surface tension and the large latent heat.

Beyond gravitation waves, our study of the confinement and chiral phase transitions can be readily employed for different models of composite dynamics for beyond standard model physics. In the future, it will be intriguing to study the impact of larger numbers of colours as well as adding dark scalars and dilatons, and therefore study the near-conformal dynamics \cite{Dietrich:2006cm}.

\begin{acknowledgments}
	MR thanks M.~Hindmarsh for helpful discussions and S.~van der Woude for correspondence on \cite{Helmboldt:2019pan}. MR acknowledges support by the Science and Technology Research Council (STFC) under the Consolidated Grant ST/T00102X/1. ZW is supported in part by the Swedish Research Council grant, contract number 2016-05996, as well as by the European Research Council (ERC) under the European Union’s Horizon 2020 research and innovation programme (grant agreement No 668679). CZ is supported by MIUR under grant number 2017L5W2PT and INFN grant STRONG and thanks the Galileo Galilei Institute for the hospitality. CZ thanks R. Garani, M. Redi and A. Tesi for discussion.
\end{acknowledgments}

\appendix

\section{Wave-function Renormalization}
\label{appendix:wfr}
The NJL model Lagrangians are constructed purely from fermion fields, as shown in \cref{eq:3F3L,eq:3G1L,eq:3S1L}. This raises the question of how to evaluate the kinetic term in the bounce action for the chiral phase transition, and more generally, also the question of how to interpret bosonic d.o.f.\ in such models. As pointed out in \cite{Helmboldt:2019pan}, the mesons should be viewed as non-propagating at tree level, but their kinetic terms can be induced by quantum corrections due to fermion loops. Therefore, unlike models in which the order parameter field is elementary, in NJL models the kinetic terms for the order parameter field need to be computed as a quantum effect. For the three gauge-fermion models of our interest, only the 3F3 model exhibits a first-order chiral phase transition, and accordingly, we need to compute the kinetic term (i.e.\ wave-function renormalization) for the $\sigma$ meson for the 3F3 case. This exercise has been done in \cite{Helmboldt:2019pan} using a 4D momentum cutoff scheme, however, some key steps in the derivation are hidden, and some subtle issues are not emphasized. Here we present the derivation for the PNJL model in the 3D momentum cutoff scheme which is the regularization scheme adopted in this work and much of the (P)NJL literature. This derivation reveals a number of subtle issues which deserve further investigation. We believe the presentation of the derivation here will also be helpful for future studies of GW signatures from a first-order chiral phase transition in other gauge-fermion models using a (P)NJL approach.

We now define
\begin{align}
	\Gbar&\equiv\Gnob_S\,,
	&
	\Gbar_D&\equiv\frac{1}{8}\Gnob_D\,,
	&
	\sigbar&\equiv-4\Gnob_S\signob\,,
	\label{eq:newnotation}
\end{align}
in the 3F3 case. The following equations are expressed in terms of the barred couplings $\Gbar,\Gbar_D$ and the barred chiral condensate $\sigbar$. These barred quantities correspond to the corresponding unbarred quantities in App.~B of \cite{Helmboldt:2019pan}. Note however we stick to the 3D momentum cutoff scheme with $\Lambda$ denoting the 3D momentum cutoff, while in \cite{Helmboldt:2019pan} $\Lambda$ denotes the 4D momentum cutoff. The 3D Euclidean bounce action can be written as
\begin{align}
	S_3=\int\! \mathrm d\Omega \,\mathrm dr\,r^2\left[\frac{Z_\sigma^{-1}}{2}\left(\frac{\mathrm d\sigbar}{\mathrm dr}\right)^2+V_{\mathrm{PNJL}}\right],
\end{align}
with $V_{\mathrm{PNJL}}$ being the PNJL grand potential, see \cref{eq:grandpotential}. If we explicitly display
the functional (and function) dependence of $S_3,\,Z_\sigma^{-1}$, and $V_{\mathrm{PNJL}}$, we have
\begin{align}
	S_3 &=S_3[\sigbar,\ell,\ell^*;T]\,,
	&
	Z_\sigma^{-1}&=Z_\sigma^{-1}(\sigbar,\ell,\ell^*,T)\,,\notag \\
	V_{\mathrm{PNJL}} &=V_{\mathrm{PNJL}}(\sigbar,\ell,\ell^*,T)\,.
	&&
\end{align}
In the PNJL model, the quark propagator is modified in the temporal background gauge field, which enters into the computation of $Z_\sigma^{-1}$. Eventually $Z_\sigma^{-1}$ also depends on the traced Polyakov loop $\ell$ as we will show. To simplify the treatment of the bounce action, we adopt the mean-field approximation for the Polyakov loop, in the sense that we set~\cite{Helmboldt:2019pan}
\begin{align}
	\ell=\ell^*=\ell_{\mathrm{min}}(\sigbar,T) \,,
\end{align}
in which $\ell_{\mathrm{min}}(\sigbar,T)$ is the value of the traced Polyakov loop that minimizes the full grand potential $V_{\mathrm{PNJL}}$ for a given value of $\sigbar$ and $T$
\begin{align}
	\frac{\partial V_{\mathrm{PNJL}}}{\partial\ell}\bigg|_{\ell=\ell_{\mathrm{min}}}=0 \,,
\end{align}
Then $Z_\sigma^{-1}$ and $V_{\mathrm{PNJL}}$ become functions of $\sigbar$ and $T$ only, and the bounce action becomes a functional of $\sigbar$ and a function of $T$: $S_3=S_3[\sigbar;T]$.

The wave-function renormalization factor $Z_\sigma^{-1}$ can be computed from the derivative of the polarization
propagator at finite temperature with respect to the spatial momentum squared~\cite{Miransky:1994vk}
\begin{align}
	Z_\sigma^{-1}=-\frac{\mathrm d\Gamma_{\sigma\sigma}(q^0,\textbf{q},\sigbar)}{\mathrm d \textbf{q}^2}\bigg|_{q^0=0,\textbf{q}^2=0} \,.
	\label{eq:zfactor1}
\end{align}
Because at finite temperature Lorentz invariance is lost, in the above equation we treat the temporal and spatial components of the external four-momentum $q$ separately. In the evaluation of $Z_\sigma^{-1}$ the derivative should be taken with respect to the spatial momentum squared in light of correspondence to the 3D bounce action. The extra minus sign in \cref{eq:zfactor1} comes about because we adopt the metric signature $(+,-,-,-)$. $\Gamma_{\sigma\sigma}(q^0,\textbf{q},\sigbar)$ is the polarization propagator for $\sigma$ at finite temperature. It is related to Feynman graphs via
\begin{align}
	\Gamma_{\sigma\sigma}(q^0,\textbf{q},\sigbar)=-i \sum\text{two-point 1PI $\sigma\text{-}\sigma$ graph}\,.
\end{align}
In the (P)NJL model, $\sigbar$ is non-propagating at classical level. To make sense of the computation of $\Gamma_{\sigma\sigma}(q^0,\textbf{q},\sigbar)$ we work in the framework of a self-consistent mean-field approximation~\cite{Kunihiro:1983ej,Hatsuda:1994pi,Helmboldt:2019pan}\footnote{In the self-consistent mean-field approximation $\sigma$ is introduced as an auxiliary field. The vacuum expectation value of $\sigma$ is also denoted $\sigma$ and it is required by the self-consistent condition to satisfy \cref{eq:newnotation}. In this appendix, $\sigma$ denotes the expectation value except in \cref{eq:lmfa1,eq:lmfa2,eq:lmfa3} and in the subscript of $\Gamma_{\sigma\sigma}$.}. The mean-field approximation Lagrangian
at finite chemical potential reads
\begin{align}
	\mL_{\mathrm{PNJL}}^{\mathrm{MFA}}=\bar{\psi}[i\slashed{\partial}-M+\gamma^0(\mu-iA_4)]\psi
	-V_{\mathrm{tree}}^{\mathrm{NJL}} \,,
	\label{eq:lmfa1}
\end{align}
in which
\begin{align}
	M&=\sigbar-\frac{\Gbar_D}{8\Gbar^2}\sigbar^2\,,
	&
	V_{\mathrm{tree}}^{\mathrm{NJL}}&=\frac{3}{8\Gbar}\sigbar^2-\frac{\Gbar_D}{16\Gbar^3}\sigbar^3\,.
	\label{eq:lmfa2}
\end{align}
One can read off the Feynman rules from the MFA PNJL Lagrangian in \cref{eq:lmfa1,eq:lmfa2}. The vertex rules for each quark colour and flavour read
\begin{align}
	\bar{\psi}\text{-}\psi\text{-}\sigbar\,\mathrm{vertex:}&\; -i \,,
	&
	\bar{\psi}\text{-}\psi\text{-}\sigbar\text{-}\sigbar\,\mathrm{vertex:}&\; \frac{i\Gbar_D}{4\Gbar^2} \,, \notag \\
	\sigbar\text{-}\sigbar\,\mathrm{vertex:}&\; -\frac{3i}{4\Gbar} \,,
	&
	\sigbar\text{-}\sigbar\text{-}\sigbar\,\mathrm{vertex:}&\; \frac{3i\Gbar_D}{8\Gbar^3} \,,
	\label{eq:lmfa3}
\end{align}
and the quark propagator reads
\begin{align}
	\mathrm{quark\,propagator:}\;\frac{i}{\slashed{k}-M+\gamma^0(\mu-iA_4)} \,.
\end{align}
Note that $iA_4$ acts as a constant imaginary chemical potential. There is no scalar propagator since scalars are non-propagating at tree level.

Because $\sigma$ has a nonzero expectation value (at zero and finite temperature),
the two-point 1PI $\sigma$-$\sigma$ graphs should contain those in which extra $\sigma$'s
are inserted, which eventually take their expectation values. We adopt the random phase approximation~\cite{Klevansky:1992qe} which implies that extra $\sigma$ insertions in the middle
of quark propagators need not be taken into account. In this approximation, their effects
are assumed to have been absorbed into the constituent quark mass. With this in mind, we arrive
at a finite set of two-point 1PI $\sigma$-$\sigma$ graphs from which we derive an expression for $\Gamma_{\sigma\sigma}(q^0,\textbf{q},\sigbar)$
\begin{widetext}
	\begin{align}
		\Gamma_{\sigma\sigma}(i\omega_E,\textbf{q},\sigbar) =-\frac{3}{4\Gbar}+\frac{3\Gbar_D}{8\Gbar^3}\sigbar
		-\left(1-\frac{\Gbar_D\sigbar}{4\Gbar^2}\right)^{\!2} N_f N_cI_S(i\omega_E,\textbf{q},\sigbar)
		+\frac{\Gbar_D}{4\Gbar^2}N_f N_cI_V(\sigbar)\,,
		\label{eq:Gammass}
	\end{align}
\end{widetext}
where $N_f=N_c=3$ and we have performed the analytic continuation $q^0=i\omega_E$, and $\omega_E$ takes value at bosonic Matsubara frequencies $2n\pi T,\,n\in\mathbb{Z}$. The two loop integrals $I_V$ and $I_S$ are defined by introducing a modified loop momentum
\begin{align}
	\bar{k}^\mu=(k^0+\mu-iA_4,\textbf{k})\,,
\end{align}
so that
\begin{align}
	I_V(\sigbar)=\frac{1}{N_c}\Tr_C\int_T\frac{\mathrm d^4 k}{i(2\pi)^4}\frac{M}{\bar{k}^2-M^2} \,,
\end{align}
and
\begin{align}
	&I_S(i\omega_E,\textbf{q},\sigbar) \notag \\
	&=\frac{1}{N_c}\Tr_C\int_T\frac{\mathrm d^4 k}{i(2\pi)^4}
	\frac{\tr[(\slashed{\bar{k}}+\slashed{q}+M)(\slashed{\bar{k}}+M)]}{[(\bar{k}+q)^2-M^2](\bar{k}^2-M^2)}\,.
\end{align}
Here $\text{tr}$ denotes the trace in Dirac space only, and $\int_T$ is defined as
\begin{align}
	\int_T\frac{\mathrm d^4 k}{i(2\pi)^4}F(k^0,\textbf{k})=
	T\sum_{n=-\infty}^{\infty}\int\!\frac{\mathrm d^3 k}{(2\pi)^3}F(i\omega_n,\textbf{k})\,,
\end{align}
with $\omega_n=(2n+1)\pi T,\,n\in\mathbb{Z}$ taking values at fermionic Matsubara frequencies.

For the computation of $Z_\sigma^{-1}$, only the term containing $I_S(i\omega_E,\textbf{q},\sigbar)$ in \cref{eq:Gammass} is relevant because only this term depends on the external momentum. By simple algebraic manipulations, $I_S(i\omega_E,\textbf{q},\sigbar)$ can be decomposed as
\begin{align}
	I_S(i\omega_E,\textbf{q},\sigbar) &=\frac{4}{N_c}\Tr_C\int_T\frac{\mathrm d^4 k}{i(2\pi)^4}\frac{1}{\bar{k}^2-M^2} \notag \\
	&\quad\,-2(q^2-4M^2)I(i\omega_E,\textbf{q},\sigbar)\,,
\end{align}
with $q^2=-\omega_E^2-\textbf{q}^2$, and $I(i\omega_E,\textbf{q},\sigbar)$ is defined as
\begin{align}
	&I(i\omega_E,\textbf{q},\sigbar) \notag \\
	&=\frac{1}{N_c}\Tr_C\int_T\frac{\mathrm d^4 k}{i(2\pi)^4}
	\frac{1}{[(\bar{k}+q)^2-M^2](\bar{k}^2-M^2)}\,.
\end{align}
With this decomposition, $Z_\sigma^{-1}$ can be computed as
\begin{align}
	Z_\sigma^{-1}=-\left(1-\frac{\Gbar_D\sigbar}{4\Gbar^2}\right)^{\!2} 2N_f N_c\left[I(0)+4M^2I'(0)\right].
\end{align}
Here $I(0)$ and $I'(0)$ are defined as
\begin{align}
	I(0)&\equiv I(i\omega_E=0,\textbf{q}=0,\sigbar)\,,
	&
	I'(0)&\equiv\frac{\mathrm dI}{\mathrm d\textbf{q}^2}\bigg|_{\omega_E=0,\textbf{q}^2=0}.
	\label{eq:I0I0p}
\end{align}
Now the crucial task is the computation of $I(i\omega_E,\textbf{q},\sigbar)$. This can be achieved by the use of partial fraction and the summation formulae~\cite{Hansen:2006ee}
\begin{align}
	\frac{1}{N_c}\Tr_C\left(\sum_n \frac{1}{i\omega_n-E_p+\mu-iA_4}\right)&=\beta f^+(E_p)\,, \notag  \\
	\frac{1}{N_c}\Tr_C\left(\sum_n \frac{1}{i\omega_n-E_p-\mu+iA_4}\right)&=\beta f^-(E_p)\,,
\end{align}
in which $\beta=\frac{1}{T},E_p=\sqrt{\textbf{p}^2+M^2}$, the summation is over fermionic Matsubara frequencies $\omega_n=(2n+1)\pi T,\,n\in\mathbb{Z}$, and the modified Fermi-Dirac distributions are defined by~\cite{Hansen:2006ee}
\begin{align}
	f^+(E_p) &=\frac{[\ell^*+2\ell e^{-\beta(E_p-\mu)}]e^{-\beta(E_p-\mu)}+e^{-3\beta(E_p-\mu)}}
	{1+3[\ell^*+\ell e^{-\beta(E_p-\mu)}]e^{-\beta(E_p-\mu)}+e^{-3\beta(E_p-\mu)}} \,, \notag \\
	f^-(E_p) &=\frac{[\ell+2\ell^* e^{-\beta(E_p+\mu)}]e^{-\beta(E_p+\mu)}+e^{-3\beta(E_p+\mu)}}
	{1+3[\ell+\ell^* e^{-\beta(E_p+\mu)}]e^{-\beta(E_p+\mu)}+e^{-3\beta(E_p+\mu)}} \,,
\end{align}
The modified Fermi-Dirac distribution satisfies the useful relation
\begin{align}
	f^-(E_p)+f^-(-E_p)=1 \,.
	\label{eq:fdproperty}
\end{align}
With the help of these summation formulae and the property \cref{eq:fdproperty} of the modified Fermi-Dirac distribution, $I(i\omega_E,\textbf{q},\sigbar)$ is computed to be (after analytic continuation back to Minkowski spacetime $i\omega_E\rightarrow\omega$)
\begin{widetext}
	\begin{align}
		I(\omega,\textbf{q},\sigbar)&=\int\frac{\mathrm d^3 p}{(2\pi)^3}\frac{1}{4E_p E_{p+q}}
		\frac{f^+(E_p)+f^-(E_p)-f^+(E_{p+q})-f^-(E_{p+q})}{\omega-E_{p+q}+E_p} \notag \\
		&\quad \,+\int\frac{\mathrm d^3 p}{(2\pi)^3}\frac{1}{4E_p E_{p+q}}\left[\frac{1-f^+(E_p)-f^-(E_{p+q})}{\omega+E_{p+q}+E_p}
		-\frac{1-f^-(E_p)-f^+(E_{p+q})}{\omega-E_{p+q}-E_p}
		\right],
	\end{align}
\end{widetext}
with $E_p=\sqrt{\textbf{p}^2+M^2}$. We take
\begin{align}
	\mu&=0\,,
	&
	\ell&=\ell^* \,,
	&
	\omega&=0\,,
\end{align}
since we work at zero chemical potential, and in the mean-field approximation $\ell=\ell^*$. Moreover according to \cref{eq:I0I0p} we only need to consider $\omega=0$. $I(0,\textbf{q},\sigbar)$ then simplifies to
\begin{align}
	I(0,\textbf{q},\sigbar)=K(0,\textbf{q},\sigbar)+L(0,\textbf{q},\sigbar)\,,
\end{align}
with
\begin{align}
	K(0,\textbf{q},\sigbar) &=\int\frac{\mathrm d^3 p}{(2\pi)^3}\frac{1}{2E_p E_{p+q}}
	\frac{f(E_{p+q})-f(E_p)}{E_{p+q}-E_p}\,, \notag \\
	L(0,\textbf{q},\sigbar) &=\int\frac{\mathrm d^3 p}{(2\pi)^3}\frac{1}{2E_p E_{p+q}}
	\frac{1-f(E_{p+q})-f(E_p)}{E_{p+q}+E_p} \,,
\end{align}
in which $f$ is the modified Fermi-Dirac distribution with $\mu=0$ and $\ell=\ell^*$
\begin{align}
	f(E_p)=\frac{\ell(1+2e^{-\beta E_p})e^{-\beta E_p}+e^{-3\beta E_p}}
	{1+3\ell(1+e^{-\beta E_p})e^{-\beta E_p}+e^{-3\beta E_p}}\,.
\end{align}
The $Z_\sigma^{-1}$ factor can then be computed as
\begin{align}
	Z_\sigma^{-1}&=\left(1-\frac{G_D\sigma}{4G^2}\right)^{\!2} 2N_f N_c\notag \\
	&\quad\,\times\left(K(0)+L(0)+4M^2[K'(0)+L'(0)]\right),
\end{align}
in which $K(0),\,L(0),\,K'(0),\,L'(0)$ are defined in the same way as $I(0),\,I'(0)$ in \cref{eq:I0I0p}. The computation of $K(0),\,L(0),\,K'(0),\,L'(0)$ gives
\begin{align}
	K(0)+L(0)&=\int\!\!\frac{\mathrm d^3 p}{(2\pi)^3}\frac{1}{4E_p^3}\left[
	1-2f(E_p)+2E_p\frac{\mathrm df}{\mathrm dE_p}\right],\notag  \\
	L'(0)&=\int\!\!\frac{\mathrm d^3 p}{(2\pi)^3}\frac{1}{16E_p^5}\left[
	6f(E_p)-2E_p\frac{\mathrm df}{\mathrm dE_p}-3\right],\notag \\
	K'(0)&=0\,,
\end{align}
which leads to the following result in the 3D cutoff scheme
\begin{widetext}
	\begin{align}
		Z_\sigma^{-1}(\sigbar,T)&=\left(1-\frac{\Gbar_D\sigbar}{4\Gbar^2}\right)^{\!2}\frac{N_f N_c}{4\pi^2}  \left(\int_0^\Lambda \!\mathrm dp\frac{p^2}{E_p^3}\left[-2f(E_p)+2E_p\frac{\mathrm df}{\mathrm dE_p}+1\right]
		+M^2\int_0^\Lambda \! \mathrm dp\frac{p^2}{E_p^5}\left[6f(E_p)-2E_p\frac{\mathrm df}{\mathrm dE_p}-3\right]
		\right).
		\label{eq:zfactorspatial}
	\end{align}
\end{widetext}
However, $Z_\sigma^{-1}$ computed via \cref{eq:zfactorspatial} becomes negative in much of the parameter space of interest. We found that this is related to the violation of Lorentz invariance in the 3D cutoff scheme at zero temperature. To illustrate the point, consider $\Gbar_D=0$ and zero temperature and chemical potential (so that all the modified Fermi-Dirac distributions become zero and drop out of the calculation). Suppose we compute $Z_\sigma^{-1}$ via the following equation
\begin{align}
	Z_\sigma^{-1}[\mathrm{temporal}]=\frac{\mathrm d\Gamma_{\sigma\sigma}(q^0,\textbf{q},\sigbar)}{\mathrm d(q^0)^2}\bigg|_{q^0=0,\textbf{q}^2=0}.
	\label{eq:zfactor2}
\end{align}
That is, the derivative is taken with respect to the square of the temporal component of the external momentum. If Lorentz invariance is maintained, \cref{eq:zfactor2} should be completely equivalent to \cref{eq:zfactor1}. However, in the 3D cutoff scheme, this is not the case. An explicit calculation at zero temperature leads to
\begin{align}
	Z_\sigma^{-1}[\mathrm{spatial}] &=\frac{9}{4\pi^2}\left(\int_0^\Lambda \mathrm dp\frac{p^4}{E_p^5}
	-2M^2\int_0^\Lambda \mathrm dp\frac{p^2}{E_p^5}\right),\notag \\
	Z_\sigma^{-1}[\mathrm{temporal}] &=\frac{9}{4\pi^2}\int_0^\Lambda \mathrm dp\frac{p^4}{E_p^5}\,,
\end{align}
in which $Z_\sigma^{-1}[\mathrm{spatial}]$ is computed via \cref{eq:zfactor1}. The difference is
\begin{align}
	\Delta Z_\sigma^{-1}&=Z_\sigma^{-1}[\mathrm{spatial}]-Z_\sigma^{-1}[\mathrm{temporal}] \notag \\
	&=-\frac{9}{2\pi^2}M^2\int_0^\Lambda \mathrm dp\frac{p^2}{E_p^5}\,.
\end{align}
Note that this difference does not vanish in the $\Lambda\rightarrow\infty$ limit
\begin{align}
	\lim_{\Lambda\rightarrow\infty}\Delta Z_\sigma^{-1}=-0.152\,.
\end{align}
So in the 3D cutoff scheme at zero temperature and chemical potential, Lorentz invariance is not restored even when we send the cutoff to infinity.

Because we indeed expect the restoration of Lorentz invariance at zero temperature and chemical potential, the above calculation reveals a limitation of the 3D cutoff scheme when applied to the evaluation of the $Z_\sigma^{-1}$ factor. A negative $Z_\sigma^{-1}$ factor signals an unphysical instability which is not acceptable in the computational framework. Motivated by the consideration of Lorentz invariance at zero temperature and chemical potential, we propose to remedy the problem by using $Z_\sigma^{-1}[\mathrm{temporal}]$ for the zero temperature contribution in $Z_\sigma^{-1}$, while the finite temperature contribution (terms that depend on $f$ or its derivative) is unchanged. This amounts to modify \cref{eq:zfactorspatial} into
\begin{widetext}
	\begin{align}
		Z_{\sigma,\mathrm{m}}^{-1}(\sigbar,T)&=\left(1-\frac{\Gbar_D\sigbar}{4\Gbar^2}\right)^{\!2}\frac{N_f N_c}{4\pi^2}  \left(\int_0^\Lambda \! \mathrm dp\frac{p^2}{E_p^3}\left[-2f(E_p)+2E_p\frac{\mathrm df}{\mathrm dE_p}+1\right]
		+M^2\int_0^\Lambda \! \mathrm  dp\frac{p^2}{E_p^5}\left[6f(E_p)-2E_p\frac{\mathrm df}{\mathrm dE_p}-1\right]
		\right),
		\label{eq:zfactormodified}
	\end{align}
\end{widetext}
with the subscript $\mathrm{m}$ in $Z_{\sigma,\mathrm{m}}^{-1}$ indicating this is the modified wave-function renormalization factor. Fortunately, this modified $Z_{\sigma,\mathrm{m}}^{-1}$ factor exhibits the expected positivity in the parameter space of interest, and thus we use it for the computation of bubble nucleation and GW signature.

Finally, as a byproduct of computing $\Gamma_{\sigma\sigma}$, here we also provide the formulae for evaluating the $\sigma$ meson mass in the 3F3, 3G1 and 3S1 models. In all three models the $\sigma$ meson mass $m_\sigma$ is determined from the equation
\begin{align}
	\Gamma_{\sigma\sigma}(\omega=m_\sigma,0,\sigbar)&=0\,,
	&
	\text{at }T&=\mu=0\,,
\end{align}
while the expression for $\Gamma_{\sigma\sigma}(\omega,0,\sigbar)$ in the three models are given by
\begin{widetext}
	\begin{align}
		\mathrm{3F3:}\quad\Gamma_{\sigma\sigma}(\omega,0,\sigbar)&=-\frac{3}{4\Gbar}+\frac{3\Gbar_D}{8\Gbar^3}\sigma
		-\frac{\Gbar_D}{16\pi^2 \Gbar^2}N_f N_cM\int_0^\Lambda \mathrm dp\frac{p^2}{E_p}
		\notag \\
		&\quad \,+\left(1-\frac{\Gbar_D\sigma}{4\Gbar^2}\right)^{\!2} \frac{N_f N_c}{\pi^2}\left[
		\int_0^\Lambda \mathrm dp\frac{p^2}{E_p}+(\omega^2-4M^2)\mathrm{P.V.}\int_0^\Lambda \mathrm dp\frac{p^2}{E_p(4E_p^2-\omega^2)}
		\right], \label{eq:Gamma1}\\
		\mathrm{3G1:}\quad\Gamma_{\sigma\sigma}(\omega,0,\sigbar)&=-\frac{1}{2\Gnob_S}
		+\frac{8}{\pi^2}\left[
		\int_0^\Lambda \mathrm dp\frac{p^2}{E_p}+(\omega^2-4M^2)\mathrm{P.V.}\int_0^\Lambda \mathrm dp\frac{p^2}{E_p(4E_p^2-\omega^2)}
		\right], \label{eq:Gamma2}\\
		\mathrm{3S1:}\quad\Gamma_{\sigma\sigma}(\omega,0,\sigbar)&=-\frac{2}{\Gnob_S}
		+\frac{6}{\pi^2}\left[
		\int_0^\Lambda \mathrm dp\frac{p^2}{E_p}+(\omega^2-4M^2)\mathrm{P.V.}\int_0^\Lambda \mathrm dp\frac{p^2}{E_p(4E_p^2-\omega^2)}
		\right]. \label{eq:Gamma3}
	\end{align}
\end{widetext}
in which we introduced the principal value integral (denoted by $\mathrm{P.V.}$ in order to get rid of the singularity in the integrand when $\omega^2=4E_p^2$, as done in \cite{Asakawa:1989bq, Kohyama:2015hix}. In fact, further simplification occurs due to the gap equations in the 3G1 and 3S1 cases,
\begin{align}
	\mathrm{3G1:}\quad&-\frac{1}{2\Gnob_S}+\frac{8}{\pi^2}\int_0^\Lambda \mathrm dp\frac{p^2}{E_p}=0\,, \notag\\
	\mathrm{3S1:}\quad&-\frac{2}{\Gnob_S}+\frac{6}{\pi^2}\int_0^\Lambda \mathrm dp\frac{p^2}{E_p}=0\,.
\end{align}
Therefore in \cref{eq:Gamma2,eq:Gamma3} the $\omega$-independent and $\sigma$-independent term cancels against the first term in the square brackets. This implies in the 3G1 and 3S1 cases the $\sigma$ meson mass simply reads
\begin{align}
	\mathrm{3G1,\,3S1:}\quad m_\sigma=2M \,.
\end{align}
This simple relation is also a consequence of the fact that we are considering the chiral limit. On the other hand in the 3F3 case due to a nonzero $G_D$, the cancellation is not perfect and $m_\sigma$ deviates from $2M$.

\section{Cosmological and Laboratory Constraints on Massive Dark Pions}
\label{appendix:constraints}
In this appendix, we show how massive dark pions can decay prior to BBN (i.e.\ with a lifetime $\tau_\pi\lesssim 1\,\rm{s}$) while being compatible with laboratory constraints. For example, consider the dimension-five Hermitian gauge-invariant effective operators
\begin{align}
	\mathcal{O}_{BE} & =(\bar{\psi}_{iL}\sigma^{\mu\nu}\psi_{jR}+\bar{\psi}_{jR}\sigma^{\mu\nu}\psi_{iL})B_{\mu\nu}\,, \notag \\
	\mathcal{O}_{BO} & =i(\bar{\psi}_{iL}\sigma^{\mu\nu}\psi_{jR}-\bar{\psi}_{jR}\sigma^{\mu\nu}\psi_{iL})B_{\mu\nu}\,,\notag  \\
	\mathcal{O}_{HE} & =(\bar{\psi}_{iL}\psi_{jR}+\bar{\psi}_{jR}\psi_{iL})H^\dagger H \,,\notag \\
	\mathcal{O}_{HO} & =i(\bar{\psi}_{iL}\psi_{jR}-\bar{\psi}_{jR}\psi_{iL})H^\dagger H \,.
\end{align}
In the above equations, $\psi$ denotes the dark quark fields, with $i,j$ being flavour indices. $B_{\mu\nu}$ is the hypercharge gauge field strength tensor, while $H$ is the SM Higgs doublet. The subscript ``$B$'' or ``$H$'' indicates the hypercharge portal or the Higgs portal respectively, while the subscript ``$E$'' or ``$O$'' indicates the CP property (even or odd). Replacing $B_{\mu\nu}$ with the dual field strength does not lead to new independent operators due to the self-duality property of the $\sigma^{\mu\nu}$ matrix. For simplicity, let us consider the case in which the dark QCD sector conserves CP. CP conservation means that we can only add the CP-even operators $\mathcal{O}_{BE}$ and $\mathcal{O}_{HE}$ to the theory. Note that $\mathcal{O}_{HE}$ can lead to the decay of CP-even (but not CP-odd) dark pions, via mixing with the SM Higgs (see ref.~\cite{Garani:2021zrr} for a lifetime estimate). In order to allow the decay of both CP-even and CP-odd dark pions, let us consider the effect of the operator $\mathcal{O}_{BE}$. With only one insertion of $\mathcal{O}_{BE}$ it is not possible to make dark pions decay as the gauge-invariant effective operator $\partial^\mu\partial^\nu\pi B_{\mu\nu}$ vanishes identically. Nevertheless, with two insertions of $\mathcal{O}_{BE}$ one may form gauge-invariant effective operators $\pi_E B_{\mu\nu}B^{\mu\nu}$ and $\pi_O B_{\mu\nu}\tilde{B}^{\mu\nu}$, with $\pi_E,\pi_O$ denoting CP-even and CP-odd dark pions respectively. The existence of these operators relies on a three-point function of composite operators in the dark QCD sector which is not protected to vanish by any symmetry. Then, both $\pi_E$ and $\pi_O$ are allowed to decay to diphoton: $\pi_E\rightarrow\gamma\gamma,\,\pi_O\rightarrow\gamma\gamma$.

With the above in mind we add the CP-conserving hypercharge portal interaction
\begin{align}
	\mathcal{L}_{BE}=\frac{c_{BE}}{\Lambda_{BE}}\mathcal{O}_{BE}\,,
\end{align}
with $c_{BE}$ a coefficient of $\mathcal{O}(1)$ and $\Lambda_{BE}$ a suppression scale. Suppose this leads to gauge-invariant effective operators
\begin{align}
	\mathcal{L}_{\rm{decay}}=a_E\pi_E B_{\mu\nu}B^{\mu\nu}+a_O\pi_O B_{\mu\nu}\tilde{B}^{\mu\nu}\,.
\end{align}
The coefficients $a_E,a_O$ can be estimated from dimensional analysis, which gives roughly
\begin{align}
	a_E,a_O\sim c_{BE}^2\frac{f_\pi}{\Lambda_{BE}^2}\,.
\end{align}
Since our analysis of phase transition and gravitational wave is performed in the massless quark limit, in the massive dark pion scenario we hope to make dark pions as light as possible (but still heavy enough to decay prior to BBN). Thus as a benchmark we take $M_\pi=10\MeV$~\footnote{Using Gell-Mann-Oakes-Renner relation, a dark quark gets bare a bare mass $\lesssim 1\MeV$. We expect its impact on the phase transition of our interest which occurs around $100\GeV$ to be small but a dedicated future study is needed to make a quantitative estimate.}. The typical dark pion decay constant relevant for our analysis can be taken as $f_\pi=100\GeV$. The dark pion lifetime can then be estimated as (for $c_{BE}\approx 1$)
\begin{align}
	\tau_\pi\sim4\pi\frac{\Lambda_{BE}^4}{f_\pi^2 M_\pi^3} \,.
\end{align}
Requiring $\tau_\pi\lesssim 1\,\rm{s}$ with the above values of dark pion mass and decay constant leads to an upper bound on the suppression scale
\begin{align}
	\Lambda_{BE}\lesssim 200\TeV \,.
\end{align}
Since we are considering a CP-conserving dark sector, we expect no constraints from electron or neutron electric dipole moment measurements or other CP-violating observables. The precision electroweak measurements and collider direct production can currently probe new physics scale of a few $\TeV$ at most, thus there exists a window of values for $\Lambda_{BE}$ which allows the decay of dark pions prior to BBN while being compatible with laboratory constraints. We also note that for $f_\pi\approx100\GeV$, dark baryons are expected to have masses around $1\TeV$ and have a relic abundance that is smaller than the required amount needed to act as dark matter (assuming zero dark baryon asymmetry)~\cite{Garani:2021zrr}. In such a case the main component of dark matter would be composed of other particles whose impact on the phase transition and gravitational wave signals can be made negligible~\footnote{The main component of dark matter can be some heavy particle outside the dark QCD sector. Its energy density scales as $\rho_{\rm{DM}}\propto a^{-3}$ ($a$ denotes the scale factor) while the radiation energy density scales as $\rho_{\rm{R}}\propto a^{-4}$ so the dark matter energy density is highly suppressed in the early Universe.}.

\bibliography{refs}
\end{document}